\documentclass[12pt]{iopart}
\usepackage{import}
\usepackage{makeidx}
\makeindex

\usepackage{graphicx}

\usepackage{iopams}  
\expandafter\let\csname equation*\endcsname\relax
\expandafter\let\csname endequation*\endcsname\relax
\usepackage{amsmath}

\usepackage{siunitx,cite}
\usepackage{color}
\usepackage[font=small,format=plain,labelfont=bf,up,textfont=normal,up]{caption}
\usepackage[breaklinks=true,colorlinks=true,linkcolor=blue,urlcolor=blue,citecolor=blue]{hyperref}

\usepackage{float}
\usepackage{pgfplots}
\usepackage{filecontents}
\usepackage{
	pst-plot,
	pst-ode
}

\usepackage{authblk}
\usepackage{amsmath}
\usepackage{commath}
\usepackage{cancel}
\usepackage{lipsum}
\usepackage{amsmath,amssymb,amsthm}
\usepackage{braket}
\usepackage{booktabs}

\newcommand{\mfaC}[1]{{\color{black}#1}}

\begin{document}

\title[Counterdiabatic driving for random-gap Landau--Zener transitions]{Counterdiabatic driving for random-gap Landau--Zener transitions}

\bigskip

\author{1}{Georgios Theologou$^1$},
\author{2,3}{Mikkel F. Andersen$^{2,3}$} and
\author{4,5}{Sandro Wimberger$^{4,5}$}
\address{$^1$\small Institute for Theoretical Physics, Universit\"at Heidelberg, Heidelberg, Germany}
\address{$^2$\small  Department of Physics, University of Otago, Dunedin, New Zealand}
\address{$^3$\small Dodd-Walls Centre for Photonic and Quantum Technologies, Dunedin, New Zealand}
\address{$^4$\small  Department of Mathematical, Physical and Computer Sciences, Parma University, Parma, Italy}
\address{$^5$\small Instituto Nazionale di Fisica Nucleare (INFN), Sezione Milano-Bicocca, Parma group, Parma, Italy}

\ead{sandromarcel.wimberger@unipr.it}
\vspace{10pt}
\begin{indented}
\item[\today]
\end{indented}

\begin{abstract}
The Landau--Zener (LZ) model describes a two-level quantum system that undergoes an avoided crossing. In the adiabatic limit, the transition probability vanishes. An auxiliary control field $H_\text{CD}$ can be reverse-engineered so that the full Hamiltonian $H_0 + H_\text{CD}$ reproduces adiabaticity for all parameter values. Our aim is to construct a single control field $H_1$ that drives an ensemble of LZ-type Hamiltonians with a distribution of energy gaps. $H_1$ works best statistically, minimizing the average transition probability. We restrict our attention to a special class of $H_1$ controls, motivated by $H_\text{CD}$. We found a systematic trade-off between instantaneous adiabaticity and the final transition probability. Certain limiting cases with a linear sweep can be treated analytically; one of them being the LZ system with Dirac $\delta(t)$ function. Comprehensive and systematic numerical simulations support and extend the analytic results.

\end{abstract}


\maketitle

\section{Introduction}
\label{sec:intro}

The Landau--Zener (LZ) system was first introduced in 1932. It describes a two-level quantum mechanical system with a linearly varying time-dependent Hamiltonian $H_0$ which undergoes an avoided crossing. Although it is time dependent, it is exactly solvable \cite{vitanov-finite,nori2}. Due to the explicit time dependence, transitions between the two eigenstates are generated, with the asymptotic transition probability being the famous LZ formula \cite{LZ-landau,LZ-zener,LZ-stuck,LZ-majorana}.

The transitions are restricted by the (quantum) adiabatic theorem \cite{born-fock}. In the adiabatic regime, the time evolution of the system follows the instantaneous eigenstate and the probability of a transition is small. The conditions under which the system remains in its initially prepared eigenstate, have both theoretical and practical significance \cite{berry-phase,farhi}. In theory, letting the system evolve slowly enough, or equivalently long enough, ensures that it follows the adiabatic trajectory, but in practice, decoherence effects and technological limitations set an upper limit to the total evolution time.

The need for fast adiabatic evolution created the field "Shortcuts to Adiabaticity" \cite{Guery, STA-2}, \mfaC{which is a subfield of the extensive and vibrant field of quantum control \cite{Dong10, DAlessandro21, Koch22}, including control of systems with imperfections or parameter variations \cite{Poggi24, Weidner24, Koch16,Li09}}. Shortcuts to Adiabaticity collectively refer to protocols that reproduce adiabatic evolution, but do not necessarily follow the adiabatic path. Relevant to this paper
is counterdiabatic Driving (CD), introduced by Demirplak and Rice \cite{demirplak1, demirplak2} and Berry \cite{berry-trans} independently. An additional control field $H_\text{CD}$ is added to the initial Hamiltonian $H_0$ such that the full system $H_0+H_\text{CD}$ drives the eigenstates of $H_0$ transitionlessly. CD driving is essentially the reverse engineering of adiabaticity. For a two-dimensional system, $H_\text{CD}$ can be found in closed form and, for low-dimensional systems, various approximation schemes \cite{wimberger-24,polk-sels} reproduce $H_\text{CD}$ sufficiently well.

The promise of fast and error-free quantum state manipulation makes CD-protocols of intense interest in quantum technology development \cite{Guery}. They have, for example, been considered for controlling superconducting qubits \cite{app1,app2} and NV centers \cite{app3}.
Theoretical investigations of CD-protocols have predominantly focused on the situation when the so-called gap-parameter (the smallest energy difference between the eigenstates of $H_0$) is fixed and known. In practical implementations, this is usually not the case, and a main motivation for choosing adiabatic transfer processes is often their robustness to variations in experimental parameters, such as the gap parameter \cite{Weitz1994, Bateman2007}. Even the paradigmatic LZ system, in which a time-dependent laser field drives two-level atoms or qubits \cite{Wubs2005}, includes a range of gaps due to variations in the intensity seen by different atoms. Some systems may also inherently include a wide range of gaps.  When analyzing two-level subspaces of a many-body spin chain, random gaps appear due to the interactions with the residual parts \cite{exper1}.
Another example is when the gap depends on the particular micro-state occupied in a thermal ensemble. For example, the coupling that drives the spin-dynamics of two thermal atoms in an optical tweezer critically depends on the particular motional state that is occupied \cite{exper2}. Any attempt to manipulate the spin-state using CD-protocols would then require that these be robust against the wide range of gaps that this leads to. Therefore, it is of both theoretical interest and practical importance to understand CD-protocols in ensembles of LZ systems with a distribution of gaps rather than a single known one. 

The main goal of this paper is the construction of a control field $H_1$ that works for an ensemble of LZ systems with a distribution of random gaps. A single control field cannot drive all the systems in the ensemble transitionlessly. Here $H_1$ is designed to work statistically, minimizing the average transition probability $\langle \mathcal{P}\rangle$. The space of time-dependent Hamiltonians, even for the two-level systems, is too large to exhaust without extra constraints. We restrict our attention to a special family of $H_1$ fields that is motivated by the standard form of $H_\text{CD}$. This leads to the definition of the Generalized LZ System, having exactly one gap parameter and one control parameter.

The generalized system is interesting in its own right, and certain limiting cases can be solved analytically. One of them is the LZ system with a Dirac $\delta(t)$ function centered at the avoided crossing. This is the limit case of an infinite control parameter, somewhat unphysical since the wavefunction becomes discontinuous in this limit. Nevertheless, it is an interesting case because it is analytically solvable and it provides insight into the optimization problem at hand. We have calculated the corresponding transition probability in closed form.

The control field that minimizes the final transition probability differs significantly from standard CD theory. In fact, by allowing the system to deviate from the adiabatic path, we achieve a lower final transition probability, on average. This is encoded as $\sigma_1$ vs $\sigma_2$  comparison, with $\sigma_i$ being the Pauli operators as control operators for our two-level problem. $\sigma_1$ control fields generally outperform  the $\sigma_2$ fields used in standard CD. This has the added advantage that $\sigma_1$ is already present in the LZ Hamiltonian, and no new control Hamiltonian needs to be engineered in an experiment. Furthermore, we find that the ideal temporal function of the control field is no longer Lorentzian.  

The structure of the paper is organized as follows: In Section \ref{sec:theory}, we give a short theoretical introduction to the LZ system and CD driving, before defining the main problem in Section \ref{sec:rg}. In Section \ref{sec:anal} we present the analytic solution for infinite control and finally, in Section \ref{sec:num}, the numerical analysis for the general case. The Appendices A, B, and C are complementary to the main text, including more details and additional calculations.

\section{Theoretical Background}
\label{sec:theory}

\subsection{The Landau--Zener problem}

In the LZ system, two asymptotically free states are swept linearly from $t= -\infty$ to $t = +\infty$ and undergo an avoided crossing at $t = 0$. The corresponding Schr\"odinger equation takes the form
\begin{equation}
    i\hbar\partial_t \ket{\psi} = (-t/T \sigma_3+a\sigma_1)\ket{\psi}
    \label{eq:LZEq}
\end{equation}
with $\sigma_i,i=1,2,3$ the Pauli matrices. The LZ formula quantifies the non-adiabatic transitions between the two eigenstates, and is given by:
\begin{equation}
    \mathcal{P}_\text{LZ} = \exp\left(-\frac{\pi T a^2}{\hbar}\right) 
    \label{eq:PLZ}
\end{equation}
In Eq. (\ref{eq:LZEq}), we can eliminate $\hbar$ and one of the parameters with the rescaling $a\rightarrow a\sqrt{T/\hbar},t\rightarrow t/\sqrt{\hbar T}$
\begin{equation}
    i\partial_t \ket{\psi} =(-t\sigma_3+a\sigma_1)\ket{\psi}
    \label{eq:LZS}
\end{equation}
and $\mathcal{P}_\text{LZ}(a) = \text{e}^{-\pi a^2}$.
The rescaled parameter $a$ characterizes the energy gap, with $2a$ being the minimum gap at the avoided crossing.

Eq. (\ref{eq:PLZ}) can be obtained by bypassing the solution of the Schr\"odinger equation \cite{LZC-1,LZC-2,LZC-3,LZD-kayanuma,LZD-rojo,LZother}. In the current work, we will need the full time evolution matrix of Eq. (\ref{eq:LZS}) \cite{vitanov-finite,nori2}.
Let $H_0\in\text{mat}(\mathbb{C},n)$ with $H_0 = H_0^\dagger$. Using the time evolution $\mathcal{U}_0$, the solution to the initial value problem $i\partial_t \ket{\psi} = H_0\ket{\psi}$ with initial condition $\ket{\psi_i}\equiv\ket{\psi(t_i)}$ is formally written as
$\ket{\psi_f}= \mathcal{U}_0(t_f,t_i)\ket{\psi_i}$. 
For $n=2$, $\mathcal{U}_0(t_f,t_i)$ is a $2\times 2$ unitary matrix that can be written in the form
\begin{equation}
    \mathcal{U}_0 = \begin{pmatrix} \mathcal{A}_0 & \mathcal{B}_0\\-\mathcal{B}_0^\ast & \mathcal{A}_0^\ast \end{pmatrix}
    \label{eq:LZU}
\end{equation}
with $\mathcal{A}_0,\mathcal{B}_0\in\mathbb{C}$ \mfaC{known as the Cayley-Klein parameters}, satisfying $\det \mathcal{U}_0=\abs{A_0}^2+\abs{B_0}^2\stackrel{!}{=}1$ \cite{qmbook-Sakurai}.
The LZ system (Eq.(\ref{eq:LZS})) is exactly solvable and $\mathcal{U}_0$ can be expressed in closed form.
Specifically, $\mathcal{A}_0,\mathcal{B}_0$ are sums of products of Parabolic Cylinder Functions \cite{vitanov-finite}.

Let $\ket{g(t)}, \ket{e(t)}$ denote the instantaneous ground and excited eigenstates respectively and $\ket{+}\equiv (1,0)^\text{T}, \ket{-}\equiv (0,1)^\text{T}$ the standard basis in $\mathbb{C}^2$.
The transition probability is defined as $\mathcal{P}_0(t_f,t_i) = \abs{\bra{e(t_f)}\mathcal{U}_0(t_f,t_i)\ket{g(t_i)}}^2$.
In the limit $t_i\rightarrow-\infty,t_f\rightarrow+\infty$, $\ket{e(+\infty)} = \ket{g(-\infty)} = \ket{-}$ and $\mathcal{P}_0 = \abs{\mathcal{A}_0}^2\sim \text{e}^{-\pi a^2}$, recovering the LZ formula as the leading-order term.


\subsection{Counterdiabatic Driving}


According to \cite{berry-trans,demirplak1,demirplak2,Guery}, we can reverse engineer a correction term $H_\text{CD}(t; a)$ such that the total Hamiltonian $H(t;a)\equiv H_0(t; a) + H_\text{CD}(t; a)$ drives the eigenstates of $H_0$ transitionlessly. \mfaC{$H_\text{CD}$ is constructed with the scope to counteract the diabatic transitions introduced by the time dependence of $H_0$.}
$H_\text{CD}$ is not unique since the eigenvectors are fixed up to complex phases \cite{wimberger-18}.
The typical choice is the one prescribed by the adiabatic theorem \cite{born-fock,qmbook-Messiah,qmbook-Sakurai}, for which 
\begin{equation}
    H_\text{CD} = i\sum_{n\neq m} \frac{\mfaC{\ket{m}\bra{m}}\partial_t H_0 \mfaC{\ket{n}\bra{n}}}{E_n-E_m},
    \label{eq:CD}
\end{equation}
with $H_0\ket{n} =E_n\ket{n}$ the instantaneous eigenstates/eigenvalues of $H_0$.
Eq.\,(\ref{eq:CD}) breaks down in the case of eigenvalue crossings or degeneracies. Moreover, for $n\geq2$ the analytic evaluation of the spectrum is a highly complicated task and instead of the exact evaluation of $H_\text{CD}$, approximation schemes are usually employed \cite{polk-sels2, wimberger-18, wimberger-24, app2}. 

With respect to the inner product $\langle A,B\rangle\equiv \tr (A^\dagger B)$, $H_\text{CD}$ obeys three orthogonality conditions \cite{wimberger-24}
\begin{equation}
    H_\text{CD}\perp \mathbf{1},\quad H_\text{CD}\perp H_0,\quad H_\text{CD}\perp \partial_t H_0
    \label{eq:CDperp}
\end{equation}
For $n=2$, Hamiltonians belong in $\text{span}_\mathbb{R}\{\mathbf{1},\sigma_{i:1,\dots,3}\}$ and the conditions above uniquely fix the direction of $H_\text{CD}$.
Thus for $n=2$, it is always possible to find $H_\text{CD}(t;a)$ and specifically for Eq.(\ref{eq:LZS}) \cite{berry-trans}, $H_\text{CD} = f_\text{CD}\sigma_2$ with
\begin{equation}
    f_\text{CD}(t;a) = \frac{1}{2}\frac{a}{t^2+a^2},
    \label{eq:fCD}
\end{equation}
which is a Lorentzian pulse centered at $t = 0$. The full Hamiltonian $H = H_0+H_\text{CD}$ then takes the form
\begin{equation}
    H(t;a) = -t\sigma_3+a\sigma_1+\frac{1}{2}\frac{a}{a^2+t^2}\sigma_2 .
    \label{eq:Htotal}
\end{equation}


\section{Random-gap LZ}
\label{sec:rg}
\subsection{Statement of the problem}

In the case of a single and fixed value $a$ for the gap, the optimal correction term is given by $H_\text{CD}$, Eq. (\ref{eq:fCD}), and the full Hamiltonian, Eq.\,(\ref{eq:Htotal}), produces a transition probability exactly $\mathcal{P} = 0$, for all times. The problem is to generalize $H_\text{CD}$ to $H_1$ which drives a distribution of gaps and minimizes the average transition probability $\langle \mathcal{P}\rangle$, generated by $H = H_0+H_1$. With $\mathcal{P}$ we denote the final (asymptotic) transition probability, in analogy to LZ formula, Eq. (\ref{eq:PLZ}).

We will assume that the gap parameter $a$ follows a Normal distribution, $a\sim \mathcal{N}(\mu,\sigma^2)$, with $\mu \equiv \langle a\rangle$ and $\sigma^2\equiv \langle (a-\mu)^2\rangle$. Our analysis is independent of the distribution, and in principle, it also works for a more general unimodal (single peak) distribution. A reasonable starting point is to use the average gap as the coupling in Eq. (\ref{eq:fCD}), i.e. $H_1 = H_\text{CD}(t;\langle a\rangle))$.
To restate the initial problem, our aim is to find a correction $H_1$ that performs better than $H_\text{CD}(t;\langle a\rangle)$, in a parameter range of $\mu,\sigma$ as large as possible.


\subsection{Default Choice} 
\label{sec:default}

\begin{figure}[!ht]
    \centering
    \includegraphics[width=1.\linewidth]{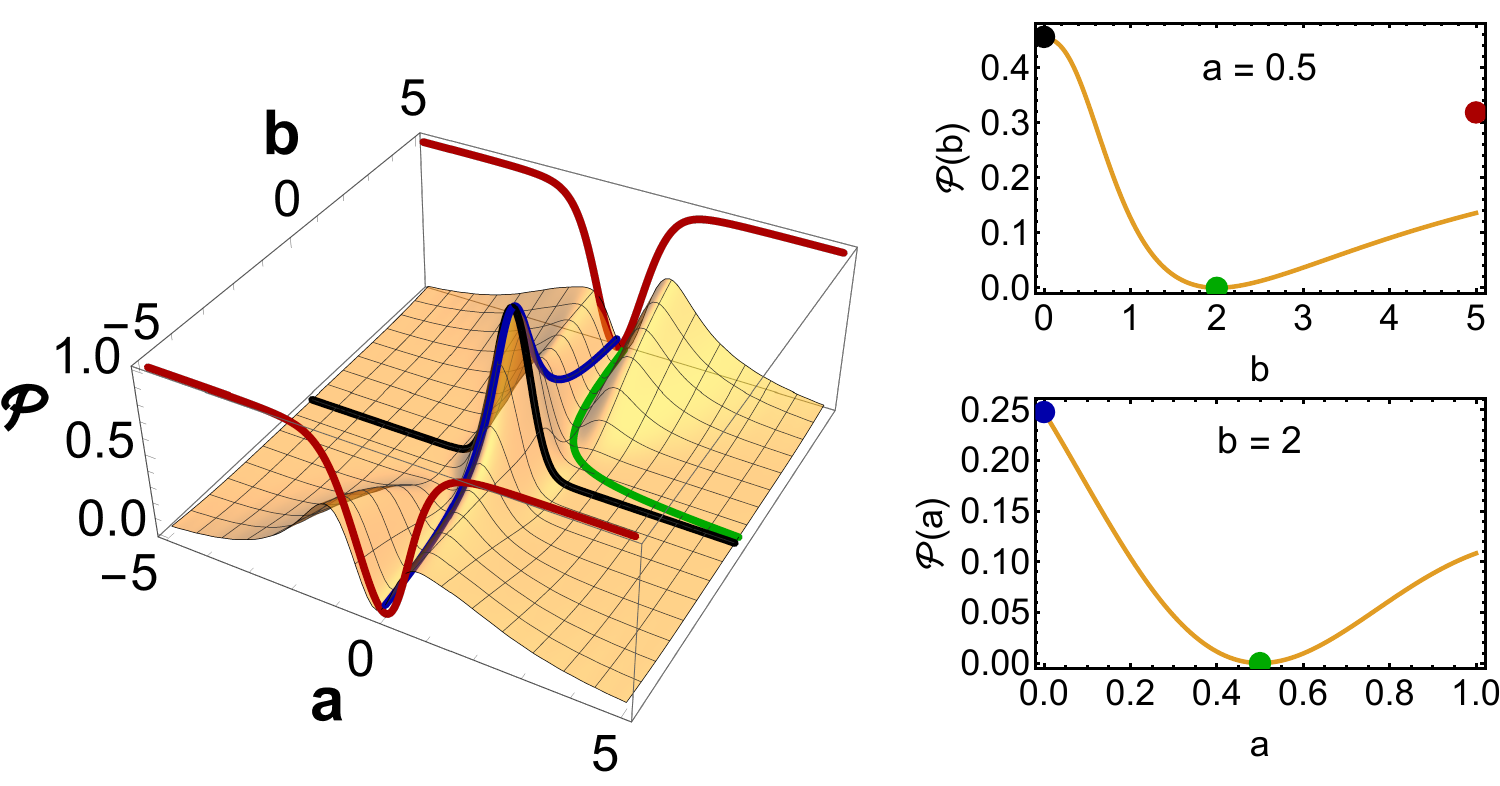}
    \caption{(Left panel) 3D graph of the transition probability generated by the Hamiltonian in Eq. (\ref{eq:HCD}). The black curve, $b = 0$, corresponds to the LZ formula, $\mathcal{P}(a,0) = \mathcal{P}_\text{LZ}(a) = \text{e}^{-\pi a^2}$. The blue curve corresponds to zero gap, $a = 0$ and the red curve corresponds to infinite control parameter, $b\rightarrow \infty$. Along the green curve, $b=1/a$, we have CD driving with $\mathcal{P}(a,1/a) = 0$. For large $a$, we can ignore the correction term and $\mathcal{P}(a,b) \stackrel{a\rightarrow \infty}{\longrightarrow}0$. Around $(0,0)$, $\mathcal{P}(a,b) \stackrel{}{\approx}\text{e}^{-\pi(a^2+b^2/4)}$. 
    $\forall a,b$, $\mathcal{P}(a,b)\in[0,1]$ and $\mathcal{P}(-a,-b) = \mathcal{P}(a,b)$, Eq. (\ref{eq:sym2}).
    Slices of $\mathcal{P}(a,b)$ are shown for constant $a$ (upper right panel) and $b$ (lower right panel), respectively.}
    \label{fig:1}
\end{figure}

As in Eq. (\ref{eq:Htotal}), we consider the Hamiltonian
\begin{equation}
    H(t;a,b) \equiv -t\sigma_3+a\sigma_1 + \frac{1}{2}\frac{b}{b^2t^2+1}\sigma_2.
    \label{eq:HCD}
\end{equation}
The control parameter $b$ and the gap coupling $a$ are now unrelated. $H$ produces transition probability $\mathcal{P}(a,b)$, effectively extending the LZ formula, $\mathcal{P}(a,0) = \mathcal{P}_\text{LZ}(a) = \text{e}^{-\pi a^2}$. $\mathcal{P}$ takes values in $[0,1]$ with $\mathcal{P}(a,b) = 0$ only along the curve
$ab-1=0$, on which we recover Eq. (\ref{eq:Htotal}). Fixing either $a$ or $b$, $\mathcal{P}$ has a global minimum at
$b_0(a)\equiv1/a$, please see the green curve in  Fig. \ref{fig:1}. 

$\mathcal{P}(a,b)$ is averaged over the gap distribution $a\sim\mathcal{N}(\mu,\sigma^2)$, resulting in a function of the control coupling $b$, $\langle\mathcal{P}\rangle (b) \equiv \langle \mathcal{P}(a,b)\rangle$, which has a minimum at $b = b^\ast$. 
We define the optimal control coupling $b^\ast$ and the corresponding minimum value $\mathcal{P}^\ast$ as
\begin{equation}
    b^\ast \equiv \underset{b>0}{\text{argmin}} \langle\mathcal{P}(a,b)\rangle_a,\quad \mathcal{P}^\ast \equiv \langle \mathcal{P}(a,b^\ast)\rangle_a
    \label{eq:pbstar}
\end{equation}

The critical values $b^\ast,\mathcal{P}^\ast$ are smooth functions of the parameters $\mu,\sigma$, and thus the optimal coupling can be expanded in power series as
\begin{equation}
    b^\ast(\mu,\sigma) \equiv \sum_{k=0}^\infty b_k(\mu) \sigma^k
    \label{eq:bstar}
\end{equation}
In the limit $\sigma\rightarrow 0$, effectively we have a single system with fixed gap $\mu = a$, and $b_0$ in Eq. (\ref{eq:bstar}) coincides with $\mathcal{P}(a,b_0(a))=0$, $b_0(\mu)=1/\mu$.

For the parameters of the distribution, in principle, we could consider arbitrary real values. However, with the only exception being section \ref{sec:anal}, we will restrict to positive $\mu>0$ and 
$\sigma\in(0,\sigma_\text{max}]$, with $\sigma_\text{max}\equiv \mu/5$. Effectively, we only allow $a>0$ since the probability of hitting a negative value 
in this parameter range is smaller than $10^{-7}$. In section \ref{sec:anal} we treat the case $\mu = 0$ separately, and we consider both positive and negative gaps. Another restriction for $0<\mu<\mu_\text{max}$ will arise after further analysis in subsequent sections. Due to the even symmetry in $\sigma$ of $\mathcal{N}(\mu,\sigma^2)$, all the odd terms in Eq.(\ref{eq:bstar}) vanish, $b_{2k+1} = 0$. Thus, for small $\mu$, $b^\ast(\mu,\sigma) \approx b_0(\mu) = 1/\mu$, we can ignore the $\sigma$ dependence.

\begin{figure}[!ht]
    \centering
    \includegraphics[width=1.0\linewidth]{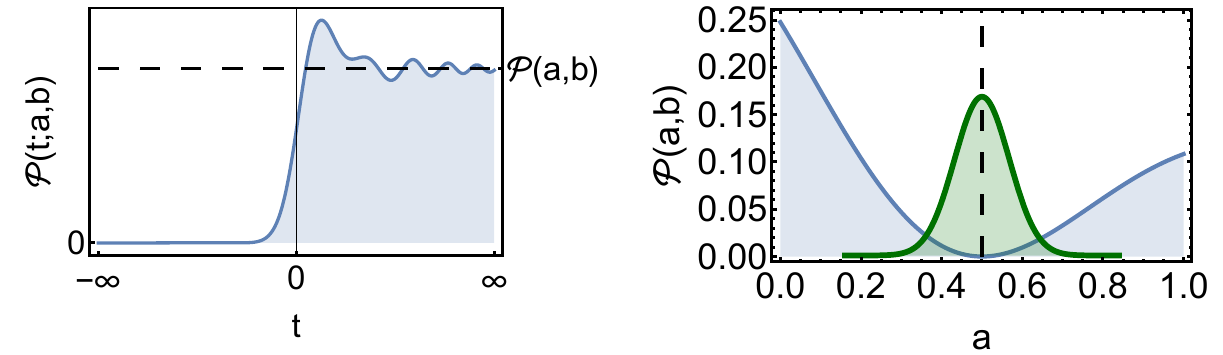}
    \caption{(Left panel) Time dependence of $\mathcal{P}(t;a,b)$. Typically, for unrelated parameters $a,b$, $\mathcal{P}(t,\cdot)$ behaves as in the original LZ system. $\mathcal{P}(a,b)$ in Fig. \ref{fig:1} denotes the final (asymptotic) value of the $\mathcal{P}(t;a,b)$. (Right panel) $\mathcal{P}$ as a function of the gap coupling $a$, for $b = 2$ (blue line). The minimum at $a=1/b =0.5$ with $\mathcal{P}(0.5,2) = 0$, corresponds to CD driving. Keeping only the the first term in Eq. (\ref{eq:bstar}) is equivalent to aligning the center of distribution with the minimum. The shaded area weighted by the Normal distribution (shown by the green curve that is not in scale), gives the optimal average transition probability $\mathcal{P}^\ast$.}
    \label{fig:2}
\end{figure}

In this approximation, $\mathcal{P}^\ast$ corresponds to the area of the $b_0-$slice of $\mathcal{P}(a,b)$, weighted by the Normal distribution, see the right panel in Fig. \ref{fig:2}.

\subsection{Generalized Landau-Zener System}

The next step is to ask whether we can do any better than the default choice, meaning that we want to find $H_1$ which performs better than
$H_\text{CD}$. As we will see in later sections, the answer is positive. We will restrict our search to a family of $H_1$ corrections which is a continuous deformation of the original $H_\text{CD}$. 

We define the Hamiltonian 
\begin{equation}
    H_\text{GLZ}(t;a,b;\varphi) \equiv -t\sigma_3+a\sigma_1+\frac{b/2}{b^2t^2+1}\sigma_\varphi
    \label{eq:GLZS}
\end{equation}
with
\begin{equation}
    \sigma_\varphi \equiv \sigma_1\cos\varphi+\sigma_2\sin\varphi,\quad \varphi\in[0,\pi/2]
\end{equation}
introducing the continuous angle parameter $\varphi$ that takes values in $\varphi\in[0,\pi/2]$. From the three orthogonality conditions that $H_\text{CD}$ satisfies, Eq. (\ref{eq:CDperp}), $H_\text{GLZ}$ violates only the second one. 

To simplify the discussion, in the remainder of this section, we will only focus on the comparison between $H_1[\varphi = 0]\propto \sigma_1$ and $H_1[\varphi = \pi/2] \propto \sigma_2$, the latter being the default case.

\subsection{Characteristic Curve}

In order for $H_1$ to be viable, the existence of a special control coupling $b_0(a)$ is essential such that $\mathcal{P}(a,b_0(a)) = 0$. Then, for a given distribution of gaps $a\sim \mathcal{N}(\mu,\sigma^2)$, we can choose as the optimal coupling (ignoring the $\sigma$ dependence) $b^\ast = b_0(\mu)$, aligning the center of the Normal distribution with the minimum, see the right panel in Fig. \ref{fig:2}.
As we increase the width $\sigma$ of $\mathcal{N}(\mu,\sigma^2)$, $b_0$ is corrected by higher order terms, Eq. (\ref{eq:bstar}), which we ignore for the most part. 

We call the roots of $\mathcal{P}(a,b;\varphi)$ collectively Characteristic Curve, defined as 
\begin{equation}
    \text{CC}[\varphi]\equiv\{(a,b_0(a;\varphi))\subset\mathbb{R}^2\ |\ \mathcal{P}(a,b_0(a;\varphi);\varphi) = 0,\ a\in\mathbb{R}\}
    \label{eq:CC}
\end{equation}
For $\varphi = \pi/2$, $b_{0}(a;\pi/2) = 1/a$, corresponding to standard CD driving, see the green curve in the left panel of Fig. \ref{fig:1}. For $\varphi \neq \pi/2$, $b_0$ must be found numerically \mfaC{since it depends on the specific form of the control field $H_1$.}

\begin{figure}[!h]
    \centering
    \includegraphics[width=1.\linewidth]{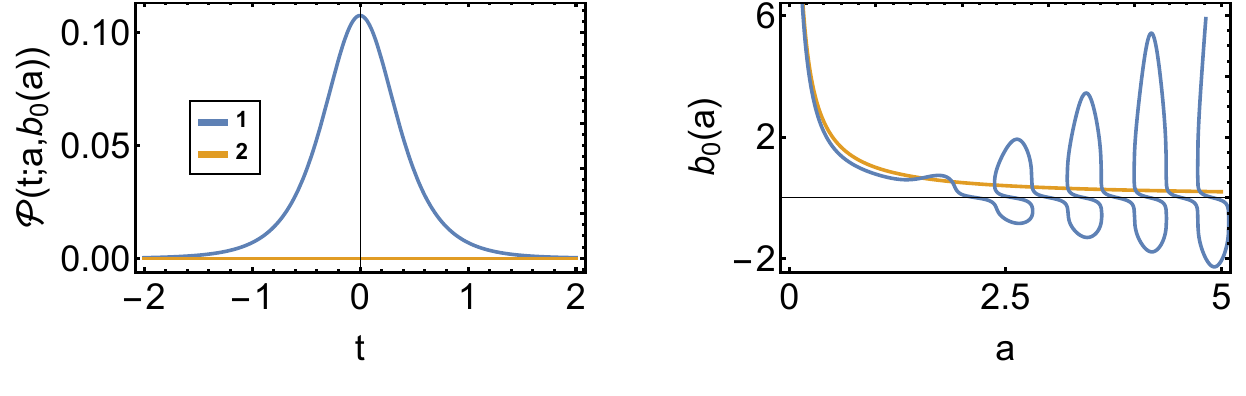}
    \caption{Comparison between $\sigma_1\ (\varphi = 0)$ (blue line) and $\sigma_2\ (\varphi = \pi/2)$ (orange line). (Left panel) Time dependence of the transition probability $\mathcal{P}(t;a,b_0(a))$ for $a= 0.5$. For the control $\sigma_2$, the counterdiabatic evolution guarantees vanishing transition probability for all times while for the control $\sigma_1$ the probability only vanishes asymptotically.
    (Right panel) Characteristic Curves, Eq. (\ref{eq:CC}). For the control $\sigma_1$, we observe irregular behavior for large $a$.}
    \label{fig:3}
\end{figure}

Regarding the properties of Eq.\,(\ref{eq:CC}), there are two crucial differences between $\varphi = \pi/2$ and the general case, see Fig. \ref{fig:3}. The first one is the time dependence of the probability at intermediate times, $\mathcal{P}(t;a,b_0(a))$. 
For $\varphi = \pi/2$, $\mathcal{P}(t;a,1/a;\pi/2) = 0,\ \forall t$, whereas for the general case, $\mathcal{P}(t;a,b_0(a;\varphi);\varphi) \stackrel{t\rightarrow +\infty}{\longrightarrow} 0$ is true only asymptotically, typically having a peak-like structure around $t = 0$. This can be seen in the left panel of Fig. \ref{fig:3}, and it presents the main compromise we make by choosing to evolve non-adiabatically.

The second difference is the global behavior. For small $a$, $b_0(a;\varphi)$ behaves similarly to $b_{0}(a;\pi/2) = 1/a$, and numerical analysis shows that $b_0$ exists, is unique, and varies smoothly with $\varphi$. For large $a$, we observe either oscillations or multivaluedness in some cases, see the right panel of Fig. \ref{fig:3}, while the origin of this behavior is not clear. To avoid this region we restrict to $0\leq\mu<2$. Otherwise, for larger $\mu$, $b_0$ should be substituted by $b^*$, see Eq. \eqref{eq:bstar}.

\subsection{Boundary Curves}

In Eq.\,(\ref{eq:GLZS}), there are the two couplings $a,b$ and the parameter $\varphi$.
Regarding the effect of the latter, it is best understood in the limit $b\rightarrow\infty$, in which $b$ can be eliminated from the problem.

\begin{figure}[!ht]
    \centering
    \includegraphics[width=0.6\linewidth]{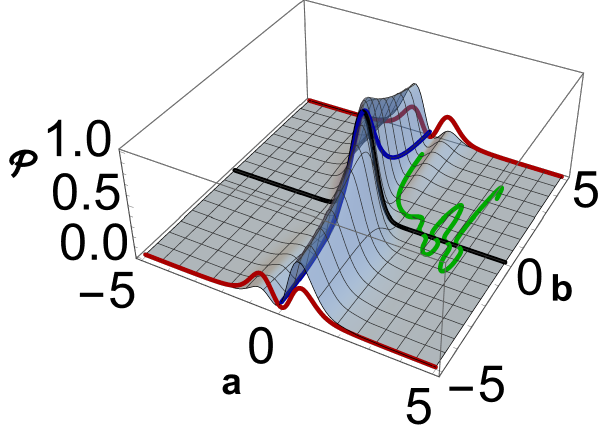}
    \caption{
        3D graph of $\mathcal{P}(a,b)$ for $\varphi = 0$, in analogy to the left panel in Fig. \ref{fig:1}. The limits $a\rightarrow0$ (blue curve) and $b\rightarrow 0$ (black curve) are unaffected by the choice of $\varphi$. The green curve is the Characteristic Curve, see the right panel in Fig. \ref{fig:3}. Comparing with Fig. \ref{fig:1}, the limit $b\rightarrow \infty$ (red curve) is better behaved, reducing the transition probability.
    }
    \label{fig:4}
\end{figure}

In Fig. \ref{fig:4}, we plot the function $\mathcal{P}(a,b)$ for $\varphi = 0$.
Comparing with $\varphi = \pi/2$, see the left panel in Fig. \ref{fig:1}, we observe that the boundary curve for $b\rightarrow\infty$ (red line), is better behaved than the one for $\varphi = \pi/2$, reducing the mass of the probability in the entire region $\{(a,b)\subset\mathbb{R}^2| a>0,b>0\}$ relevant to our problem. 

In the next section, we treat the limit $b\rightarrow \infty$ analytically and show that $\varphi = 0$ is the optimal value in this limit.

\section{Landau--Zener problem with Dirac-$\delta$ coupling}
\label{sec:anal}

In the limit $b\rightarrow\infty$, the Schr\"odinger equation for Eq.\,(\ref{eq:GLZS}) can be solved analytically and the resulting transition probability
$\mathcal{P}_\infty(a;\varphi)\equiv\lim_{b\rightarrow\infty}\mathcal{P}(a,b;\varphi)$ can be found in closed form (Eq.\,(\ref{eq:redprob})). Instances of these functions have already appeared as red curves in Fig. \ref{fig:1} $(\varphi = \pi/2)$ and Fig. \ref{fig:4} ($\varphi = 0$). Eq.\,(\ref{eq:redprob}) is a special case of Eq.\,(\ref{eq:redProb2}), which is derived in \ref{ap:red}, including more details. The main result is that the value $\varphi = 0$, minimizes $\mathcal{P}_\infty(a;\varphi)$ for all values of the gap parameter $a$. With regard to the random gap problem introduced in the previous section, the limit $b\rightarrow \infty$ corresponds to $\mu \rightarrow 0$, with $b_0(\mu;\varphi)\stackrel{\mu\rightarrow 0}{\longrightarrow} \infty$ independently of $\varphi$, see the right panel in Fig. \ref{fig:3}. The conclusion is that in the limit $b\rightarrow \infty$, $\varphi = 0$ minimizes the average transition probability for any probability distribution with an average $\mu=0$.

\subsection{Preliminaries}

For $b\rightarrow \infty$, the Schr\"odinger equation for the generalized Hamiltonian, Eq. (\ref{eq:GLZS}), takes the form
\begin{equation}
    i\partial_t\ket{\psi} = \bigg[-t\sigma_3+a\sigma_1+\frac{\pi}{2}\delta(t)\sigma_\varphi\bigg]\ket{\psi}
    \label{eq:redMain}
\end{equation}
with $\delta(t)$ the Dirac-delta function. Eq. (\ref{eq:redMain}) is first order in time. Due to the $\delta(t)$ in the r.h.s, the solution must have a jump discontinuity at $t= 0$, with $\ket{\psi(0^+)} = \mathcal{K} \ket{\psi(0^-)}$. 
The matrix $\mathcal{K}$ is fixed by Eq. (\ref{eq:redMain}) with $\mathcal{K} = \text{e}^{-i\frac{\pi}{2}\sigma_\varphi}$. Using the properties of the Pauli matrices, the matrix exponential can be expressed in closed form, see \ref{ap:DC},
\begin{equation}
    \mathcal{K}(\varphi) =  -i\sigma_\varphi .
    \label{eq:DC}
\end{equation}

The solution to Eq. (\ref{eq:redMain}) reduces to the LZ system (Eq. (\ref{eq:LZS})) by splitting the time evolution $\mathcal{U}_\infty(t_f,t_i;a,\varphi) = \mathcal{U}_0(t_f,0;a)\mathcal{K}(\varphi)\mathcal{U}_0(0,t_i;a)$ and taking the limits $t_f = -t_i,t_f\rightarrow +\infty$, leading to $\mathcal{P}_\infty = \abs{\bra{-}\mathcal{U}_\infty\ket{-}}^2= \abs{\mathcal{A}_\infty}^2$, with $\mathcal{A}_\infty$ being the diagonal elements of $\mathcal{U}_\infty$. 

\mfaC{In Eq. \eqref{eq:redMain} the Dirac-$\delta(t)$ function can be seen as the limit of an instantaneous pulse with infinite intensity. An alternative, more intuitive interpretation is the following. For the two level system, the physical states can be uniquely mapped onto the surface of the three dimensional sphere, also known as Bloch Sphere \cite{qmbook-NielsenChuang}. Geometrically, the net effect of the $H_1 = \frac{b/2}{b^2t^2+1}\sigma_\varphi$ is a $\pi$-rotation around the axis $\hat{n}_\varphi = (\cos\varphi,\sin\varphi,0)$, regardless of $b$, since $i\partial_t\ket{\psi} = H_1 \ket{\psi} \Rightarrow$ $\ket{\psi(+\infty)} = \text{e}^{-i \frac{\pi}{2}\sigma_\varphi} \ket{\psi(-\infty)}$.
For finite $b$, the rotation is applied gradually thought the time evolution $t\in(-\infty,+\infty)$. With this representation in mind, in the limit $b\rightarrow\infty$, the effect of $H_1 \stackrel{b\rightarrow\infty}{\longrightarrow} \frac{\pi}{2}\delta(t)\sigma_\varphi$ is an instantaneous $\pi$-rotation in the Bloch sphere.
}

\subsection{Solution with Dirac-$\delta$ coupling}

The case $a = 0$ is trivial and is treated first. Eq. (\ref{eq:redMain}) reduces to
\begin{equation}
    i\partial_t\ket{\psi} = \bigg[-t\sigma_3 + \frac{\pi}{2}\delta(t) \sigma_1\bigg]\ket{\psi}.
    \label{eq:blueDelta}
\end{equation}
For $t_f>0$, integrating Eq. (\ref{eq:blueDelta}) in the symmetric interval $[-t_f,t_f]$ with $\ket{\psi(-t_f)} = \ket{-}$ and taking into account the discontinuity, Eq. (\ref{eq:DC}), we get $ \ket{\psi(t_f)} = -i\sigma_\varphi \ket{\psi(-t_f)} = -i \text{e}^{-i\varphi}\ket{+}$ leading to $\mathcal{P}_\infty(0;\varphi) = 0$, $\forall \varphi$, by taking the limit $t_f\rightarrow\infty$.

For the general case with $a\neq 0$, the full time evolution matrix to Eq. (\ref{eq:redMain}) is 
\begin{equation}
    \mathcal{U}_\infty(t_f,t_i;a,\varphi) = \mathcal{U}_0(t_f,0;a) \mathcal{K}(\varphi)\mathcal{U}_0(0,t_i;a)
    \label{eq:redU}
\end{equation}
Expanding the matrices in Eq. (\ref{eq:redU}) we obtain (suppressing the $a$ dependence)
\begin{equation}
    \mathcal{A}_\infty(t_f,t_i) = -i\bigg[\mathcal{A}_0(0,t_i)\mathcal{B}_0(t_f,0)\text{e}^{i\varphi} -
        \mathcal{A}_0(t_f,0)\mathcal{B}^\ast_0(0,t_i)\text{e}^{-i\varphi}\bigg],
        \label{eq:redA}
\end{equation}
with $\mathcal{A}_0,\mathcal{B}_0$ given by Eqs. (\ref{eq:LZSA}) and (\ref{eq:LZSB}).
In order to exploit the time symmetry, Eq. (\ref{eq:sym1}), we consider symmetric integration around $t = 0$, with  $t_f = -t_i$. After algebraic manipulations, we have
\begin{align}
    \mathcal{A}_\infty(t_f,-t_f)
        &= -i\bigg[\mathcal{A}_0(0,-t_f)\mathcal{B}_0(t_f,0)\text{e}^{i\varphi}
            - \mathcal{A}(t_f,0)\mathcal{B}^\ast_0(0,-t_f)\text{e}^{-i\varphi}\bigg]\nonumber\\
        &= -i\bigg[\mathcal{A}_0(0,-t_f)\mathcal{B}_0(0,-t_f)\text{e}^{i\varphi}
            - \mathcal{A}^\ast (0,-t_f)\mathcal{B}^\ast_0(0,-t_f)\text{e}^{-i\varphi}\bigg]\nonumber\\
        &= 2\text{Im} (\underbrace{\mathcal{A}_0(0,-t_f)\mathcal{B}_0(0,-t_f)}_{\mathcal{C}_0}\text{e}^{i\varphi})
        \label{eq:redA2}
\end{align}
Keeping only leading order terms in the limit $t_f\rightarrow+\infty$, $\mathcal{C}_0$ in Eq. (\ref{eq:redA2}) takes the form
\begin{equation}
    \mathcal{C}_0 \sim\frac{\Gamma(1-\nu)}{a\sqrt{\pi}}D_\nu^2(0)(\text{e}^{-\pi a^2/2}-1)e^{i\frac{\pi}{4}},
    \label{eq:redC}
\end{equation}
with $\nu(a)\equiv i a^2/2$ and $D_\nu(z)$ the Parabolic Cylinder Functions. 
Further simplification can be made using the formula $D_\nu(0) =2^{\frac{\nu}{2}}\sqrt{\pi}/\Gamma\left(\frac{1-\nu}{2}\right)$, \cite[Chapter 19]{abramowitz} and properties of the Gamma function, see \ref{ap:special}. The final form for $\mathcal{P}_\infty = \abs{\mathcal{A}_\infty}^2$ is
\begin{equation}
    \mathcal{P}_\infty(a;\varphi) = (1-\text{e}^{-\pi a^2})\cos^2(\chi(a)-\varphi),
    \label{eq:redprob}
\end{equation}
with
\begin{equation}
    \chi(a) \equiv \frac{\pi}{4} + \arg\Gamma\left(\frac{1-\nu(a)}{2}\right)-\arg\Gamma\left(\frac{2-\nu(a)}{2}\right),\quad \nu(a) \equiv i \frac{a^2}{2}.
    \label{eq:chi}
\end{equation}

$\mathcal{P}_\infty(a;\varphi)$ is the complement probability of the LZ formula $\mathcal{P}_0(a) = \text{e}^{-\pi a^2}$, weighted by the factor $\cos^2(\chi(a)-\varphi)$. $\chi(a)$ satisfies $\chi(-a) = \chi(a)$ with $\chi(0) = \pi/4$ and for $a>0$ increases from $\pi/4$ to $\pi/2$. It is related to the time derivatives of the transition probability of the LZ system at $t = 0$ and the so-called jump time, introduced in \cite{vitanov-time}. 
Its asymptotic expansions can be found in \cite{vitanov-time}, from which, in combination with Eq. (\ref{eq:redprob}), we obtain
\begin{equation}
    \mathcal{P}_\infty(a;\varphi) \stackrel{a\rightarrow 0}{\sim}
        \pi a^2 \cos^2(\varphi-\tfrac{\pi}{4}) - \frac{\pi a^4}{2}\left[\ln2 \cos 2\varphi+\pi \cos^2(\varphi-\tfrac{\pi}{4}) \right]
            +\mathcal{O}(a^6)
    \label{eq:redAs0}
\end{equation}
\begin{equation}
    \mathcal{P}_\infty(a;\varphi)-\sin^2\varphi \stackrel{a\rightarrow\infty}{\sim} \frac{1}{2a^2}\sin2\varphi + \frac{1}{4a^4}\cos2\varphi + \mathcal{O}(1/a^6)
    \label{eq:redAsInf}
\end{equation}

In Eq. (\ref{eq:redAs0}), the first non-vanishing term depends only on $\abs{\tfrac{\pi}{4}-\varphi}$ and it takes its minimum at $\varphi = 0,\pi/2$. 
The next order term separates the two values, making $\mathcal{P}_\infty(a;0)$ minimal. 
In the other limit, Eq. (\ref{eq:redAsInf}), $\mathcal{P}_\infty$ approaches $\sin^2\varphi$ as $1/a^2$, except for $\varphi = 0,\pi/2 
\Rightarrow \sin2\varphi = 0$, for which the first order vanishes. Only for $\varphi = 0$, 
$\mathcal{P}_\infty(a;0)\stackrel{a\rightarrow \infty}{\longrightarrow}0$.

From the observations above, we conclude that the parameter $\varphi = 0$ is the optimal value for all $a$, minimizing the corresponding probability, i.e.,
\begin{equation}
    \mathcal{P}_\infty(a;0) = \min_{\varphi\in[0,\pi/2]}\{\mathcal{P}_\infty(a;\varphi)\},\quad a\in\mathbb{R}.
    \label{eq:redMin}
\end{equation}

\begin{figure}[!tb]
    \centering
    \includegraphics[width=1.\linewidth]{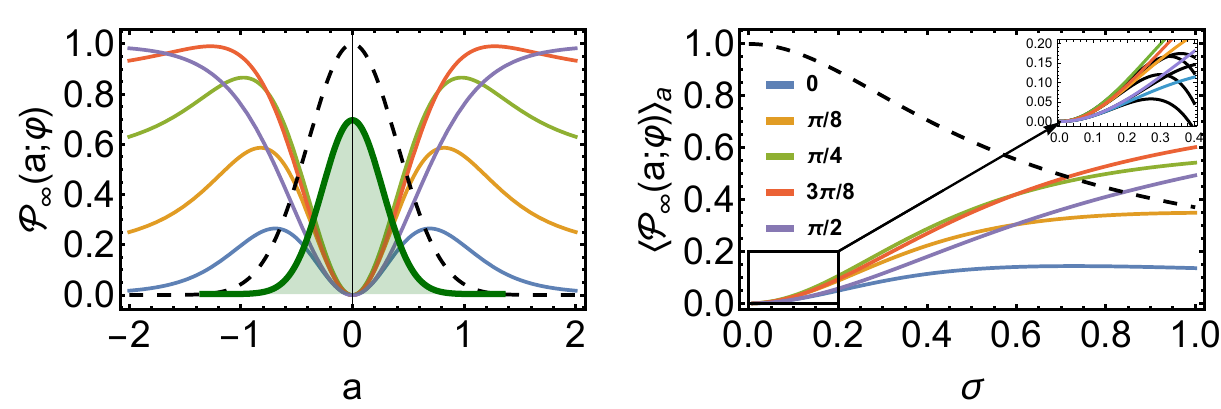}
    \caption{(Left panel) Plot of Eq. (\ref{eq:redprob}) as a function of the gap coupling $a$, for different values of $\varphi$, superimposed with the Normal distribution, centered at $\mu = 0$ (not in scale). (Right panel) Average probability over $a\sim\mathcal{N}(0,\sigma^2)$ as a function of $\sigma$. Colors are the same in both figures according to the legend in the right panel. \mfaC{The dashed lines correspond to the LZ formula $\mathcal{P}_\text{LZ}(a) = \text{e}^{-\pi a^2}$, Eq. \eqref{eq:PLZ} (left panel), and $\langle\mathcal{P}_\text{LZ} \rangle(0,\sigma) = 1/\sqrt{1+2\pi\sigma^2}$, Eq. \eqref{eq:PLZav} (right panel). The inset shows a zoom around the origin with the black lines calculated via Eq. \eqref{eq:redAs0}.}
    } 
    \label{fig:5}
\end{figure}

Returning to the random gap problem, for $a\sim \mathcal{N}(0,\sigma^2)$, $b_0(0;\varphi)\rightarrow\infty$, which is independent of $\varphi$ in this limit, Fig. \ref{fig:3}, is chosen such that $\mathcal{P}(0,\infty) = 0$. As a consequence of Eq. (\ref{eq:redMin}), $\varphi = 0$ also minimizes the average probability. To quantify the advantage, in Fig. \ref{fig:5} (right panel), we plot the average as a function of the standard deviation $\sigma$ with $a\sim \mathcal{N}(0,\sigma^2)$. We observe significant improvement for $\varphi = 0 $. Eventually, for large $\sigma$, all the curves converge to zero due to the normalization factor $1/\sqrt{2\pi\sigma^2}$ in the Normal distribution. \mfaC{Equation \eqref{eq:redAs0}, with the substitutions $\langle a^2\rangle = \sigma^2,\ \langle a^4\rangle = 3\sigma^4$, provides a good approximation for $\sigma\in(0,0.2)$, see inset in the right panel of Fig. \ref{fig:5}.}

\mfaC{Lastly, we would like to compare our results with the bare Hamiltonian $H_0$, Eq. \eqref{eq:LZS}. The asymptotic transition probability is given by $\mathcal{P}_\text{LZ}(a) = \text{e}^{-\pi a^2}$, Eq. \eqref{eq:PLZ}. For $a\sim \mathcal{N}(0,\sigma^2)$ the average value is
$\langle \mathcal{P}_\text{LZ}\rangle = \frac{1}{\sqrt{1+2\pi \sigma^2}}$, Eq. \eqref{eq:PLZav} for $\mu = 0$. The corresponding curve is shown in the right panel of Fig. \ref{fig:5} (black dashed line). It intersects the $\varphi = \pi/2$ line at $\sigma^\ast_{\pi/2} \approx 0.84$, whereas for $\varphi = 0$ they curves do not intersect at all, meaning that the inclusion of the external field $H_1[\varphi = 0]$ is advantageous to the bare Hamiltonian $H_0$ for arbitrarily large $\sigma$ and asymptotically gives an improvement of a factor of the order 2.}

What has been shown in this section remains valid for any unimodal distribution with a zero mean, not just for the Normal distribution we are focusing on here. The goal of the next section is to investigate numerically whether the results of this section extend to finite mean values $\mu>0$ and thus finite-control couplings $b$.

\section{Numerical simulation for the general case}
\label{sec:num}

The main result of the previous section is that $\varphi = 0$ minimizes the average probability in the limit of an infinite control term, $b\rightarrow\infty$, which is optimal for the distribution $a\sim\mathcal{N}(0,\sigma^2)$.
The purpose of the current section is to investigate the extension to $\mu>0$ ($b$ finite) which can only be done numerically.

Once again, we will only address the two edge cases $\sigma_1, (\varphi = 0)$ and $\sigma_2,\ (\varphi = \pi/2)$. The main finding is that $\sigma_1$ outperforms $\sigma_2$ in a wide range of the parameter $\mu,\sigma$, but at the cost of instantaneous adiabaticity, see Fig. \ref{fig:7} below.

First, we establish the result numerically. Then, we perform a stability analysis of experimental errors. Finally, we modify our model, Eq. (\ref{eq:GLZS}), by either changing the shape of the pulse function in Eq. (\ref{eq:GLZS}) or the sweep function.
Although our analysis in those two subsections is not exhaustive, the results follow the same theme, showing a small but systematic advantage of $\sigma_1$ over $\sigma_2$ across a broad range of parameters.

Throughout this section, for the distribution of gaps, we assume a Normal distribution $a\sim\mathcal{N}(\mu,\sigma^2)$ with $\sigma\in(0,\sigma_\text{max}]$, $\sigma_\text{max} \equiv \mu/5$. Instead of Eq. (\ref{eq:LZS}), we use its rescaled version
\begin{equation}
    i\partial_s \ket{\psi} = \bigg[T(-\lambda(s) \sigma_3 +a\sigma_1) + \partial_s\lambda\frac{b/2}{b^2\lambda(s)^2+1}\sigma_\varphi\bigg]\ket{\psi},\ s\in[0,1],
    \label{eq:LZST}
\end{equation}
with the same solutions for large $T$. For all the simulations, apart from the last subsection, $\lambda(s) = \lambda_0(s-\tfrac12),\ T = \lambda_0 = 10$. Unless stated otherwise, we make the approximation $b^\ast = b_0$, see Eq. (\ref{eq:bstar}), ignoring higher-order terms. By default, for the numerical averages, \mfaC{we use 1000 simulations. The only exception is Fig. \ref{fig:6}, in which we compare different sample sizes.}

\subsection{$\sigma_1$ vs. $\sigma_2$ control}
\label{sec:s1-vs-s2}

\begin{figure}[!ht]
    \centering
    \includegraphics[width=1.\linewidth]{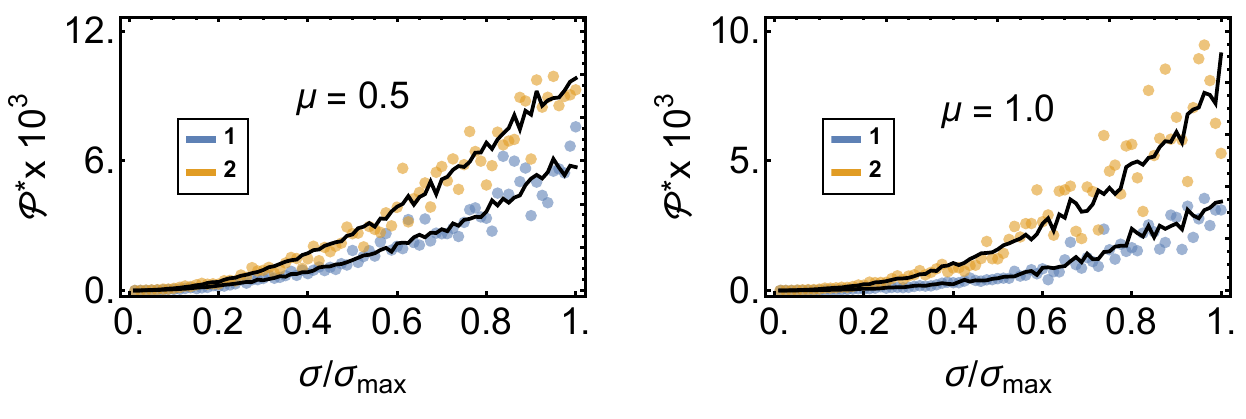}
    \caption{$\mathcal{P}^\ast$ as a function of $\sigma$ for $\mu = 0.5,\ \sigma_\text{max} = 0.1$ (left panel) and $\mu=1.0,\ \sigma_\text{max} = 0.2$ (right panel). The two curves correspond to $\varphi = 0$ ($\sigma_1$, blue line) and $\varphi = \pi/2$ ($\sigma_2$, orange line). Points are averaged over 100 and black lines over 1000 random $a$ values. \mfaC{Comparing with the average of the bare Hamiltonian $H_0$, Eq. \eqref{eq:PLZav}, the corresponding values at $\sigma = \sigma_\text{max}$ are $\langle\mathcal{P}_\text{LZ} \rangle \approx 463 \times 10^{-3}$ (left panel) and $\langle\mathcal{P}_\text{LZ} \rangle \approx 73 \times 10^{-3}$ (right panel).}
    }
    \label{fig:6}
\end{figure}

In Fig. \ref{fig:6}, we see the typical behavior of the mean transition probability $\mathcal{P}^\ast$ for two distinct values of $\mu$. For small width $\sigma$ of the distribution of $a$, the two curves and their distance increase with $\sigma$. \mfaC{As expected, the random fluctuations diminish with increasing number of sample points.}

It is easy to see that all curves in Fig. \ref{fig:6} should be monotonically increasing for small $\sigma$ and any deviations from monotonicity are due to statistical fluctuations in our numerical simulations. By definition,
$\langle\mathcal{P}\rangle = \frac{1}{\sqrt{2\pi\sigma^2}}\int_\mathbb{R}\text{d}a\ \text{e}^{-(a-\mu)^2/(2\sigma^2)}\mathcal{P}(a,b)$.
Expanding around $a = \mu$ and $b^\ast = b_0$, we obtain $\mathcal{P}^\ast(\mu,\sigma)\stackrel{\sigma\rightarrow 0}{\sim}\frac{1}{2}\mathcal{P}''(\mu,b_0(\mu))\sigma^2>0$, with $\mathcal{P}''(\mu,b_0(\mu))>0$ because of the minimum, see the right panel in Fig. \ref{fig:1}.

\begin{figure}[!ht]
    \centering
    \includegraphics[width=1.\linewidth]{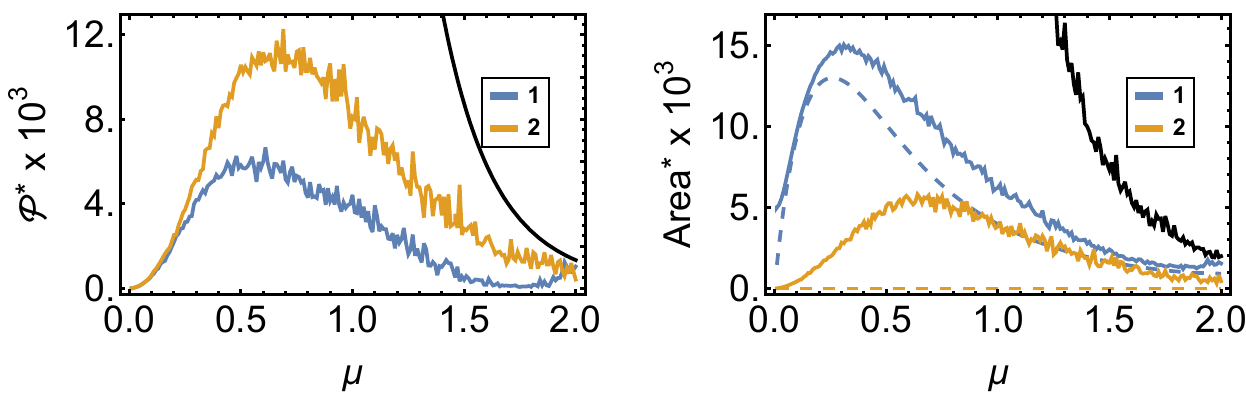}
    \caption{$\mathcal{P}^\ast$ (left panel) and average area in the time domain (right panel) as functions of $\mu$, for a distribution $a\sim\mathcal{N}(\mu,\mu^2/25)$. Dashed lines on the right panel correspond to single gap ($\sigma \rightarrow 0$). For $\sigma_2$, the area is theoretically exactly zero while for $\sigma_1$ there is a deviation from zero due to peak around $t = 0$, Fig. \ref{fig:3} (left panel). \mfaC{Black lines corresponds to the evolution of the free Hamiltonian $H_0$, Eq. \eqref{eq:PLZav} with $\sigma = \mu/5$ for the left panel and the numerically calculated average for the right panel.}
    }
    \label{fig:7}
\end{figure}

Next we want to quantify the difference between $\sigma_1$ and $\sigma_2$ for different values of $\mu\in(0,2)$, which is done in Fig. \ref{fig:7}. 
Given $\mu$, we set $\sigma = \sigma_\text{max} = \mu/5$, such that the separation between the two curves is maximal, see Fig. (\ref{fig:6}). 
On the left panel we plot the average probability $\mathcal{P}^\ast$. In combination with Fig. \ref{fig:6} it shows that $\sigma_1$ is better than $\sigma_2$ in the range of parameters $\mu\in(0,2),\ \sigma\in (0,\mu/5)$.
There must be a peak in the curves shown in Fig. \ref{fig:7} since in the limit of vanishing $\mu$, with the restriction on $\sigma$, the transition probability can be made small by a single control value $b$. In the other limit of large $\mu$, the transition probability goes to zero because of normalization, see \ref{ap:limits}.

The main compromise we are making by using a correction term other than $H_\text{CD}$, is that the time evolution is no longer counterdiabatic during the entire evolution. For a single realization of the gap $a$, $\mathcal{P}(t;a,b_0(a))$ only vanishes asymptotically, typically having a peak around $t =0$, see the left panel in Fig. \ref{fig:3}.
In the right panel of Fig. \ref{fig:7}, we compute the average area (in the time domain), of the transition probability from Fig. \ref{fig:2} (left panel), which measures the deviation from adiabaticity, on average.
For $\sigma_2$, the single control field $H_\text{CD}$ cannot drive the full ensemble of systems transitionlessly. Nevertheless, the average deviation from zero is small at intermediate times. The adiabatic theorem ensures that $\mathcal{P}(t;a,b_0(a)) \stackrel{a\rightarrow \infty}{\longrightarrow} 0, \forall t$, and therefore both curves in the right panel of Fig. \ref{fig:7} go to zero for large $\mu$.

\subsection{Stability analysis}

In the previous section, we compared between the $\sigma_1$ and the $\sigma_2$ control. The advantage of $\sigma_1$ over $\sigma_2$ is that for a wide parameter range $\mathcal{P}^\ast[\sigma_1] < \mathcal{P}^\ast[\sigma_2]$, see Fig. \ref{fig:6} and the left panel of Fig. \ref{fig:7}. The compromise for this advantage is that $\mathcal{P}^\ast$ deviates from the adiabatic path noticeably for intermediate times, see the right panel in Fig. \ref{fig:7}.

The next step is to analyze the stability of the two different controls. For this, we investigate the effect of different types of errors by introducing an imperfection parameter $\varepsilon$ in the correction term $H_1$, see Tab.\,\ref{tab:error}. The first type of error is the one that is usually taken into consideration in an experimental setup. The parameter $\varepsilon$ affects both the peak and the area of the pulse.
In the two other cases, we have made the appropriate modifications such that either the area or the height of the pulse is kept constant, up to first order in $\varepsilon$. This would require extra control of the experimental setup and, in principle, is harder to implement in practice.

\begin{table}[!h]
\def\arraystretch{1.2}
    \centering
    \begin{tabular}{r|c|c|c}
    \toprule
    {}&Function & Area & Peak\\
    \midrule
     (1)  &  $ (1+\varepsilon)\frac{b/2}{b^2t^2+1}$ &
            $\frac{\pi}{2}(1+\varepsilon)$ &
               $(1+\varepsilon)\frac{b}{2}$ \\
    (2)   & $ \frac{b/2}{b^2 t^2(1-\varepsilon)^2+1}$ &
                $\frac{\pi}{2}\frac{1}{1-\varepsilon}=\frac{\pi}{2}(1+\varepsilon+\dots)$ &
                    $\frac{b}{2}$ \\
    (3)   & $ (1+\varepsilon)\frac{b/2}{b^2t^2(1-\varepsilon)^2+1}$ &
            $\frac{\pi}{2}\frac{1+\varepsilon}{1-\varepsilon} = \frac{\pi}{2}(1-\varepsilon^2+\dots)$ &
               $(1+\varepsilon)\frac{b}{2}$ \\
    \bottomrule
    \end{tabular}
    \caption{Different types of imperfection in the correction term $H_1$. (1) The error affects both area and peak. (2) Fixing the peak and (3) fixing the area up to first order in $\varepsilon$.}
    \label{tab:error}
\end{table}

\begin{figure}[!bth]
    \centering
    \includegraphics[width=1.\linewidth]{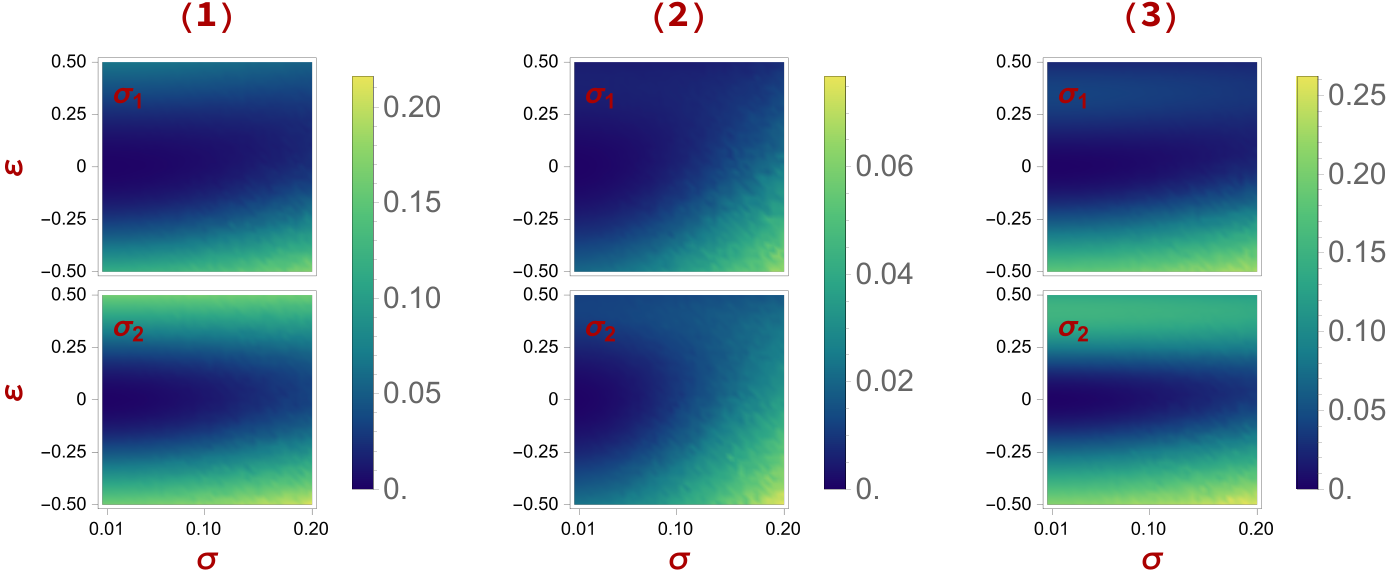}
    \caption{Heatmap of $\mathcal{P}^\ast$ for the different types of errors from Tab. \ref{tab:error}  and gap distribution $a\sim\mathcal{N}(\mu,\sigma^2)$, $\mu=0.5$. Blue regions are desired, corresponding to low average transition probability. The top row corresponds to $\sigma_1$ and the bottom row to $\sigma_2$. Different columns correspond to different types of errors. In each subplot, in the horizontal axis we vary $\sigma$ and in the vertical axis the error parameter $\varepsilon$, see Tab. \ref{tab:error}. Fixing the peak, type (2) in Tab. \ref{tab:error}, improves the accuracy approximately by a factor of three. The qualitative behavior between $\sigma_1$ and $\sigma_2$ is the same.}
    \label{fig:8}
\end{figure}

In Fig. \ref{fig:8} we plot the heatmap of $\mathcal{P}^\ast$. On the horizontal axis, we vary the standard deviation of the distribution of gaps $a\sim \mathcal{N}(\mu,\sigma^2)$ and, on the vertical axis, we vary the imperfection parameter $\varepsilon$. Our numerical analysis indicates that the stability of the two cases, $\sigma_1$ and $\sigma_2$, is qualitatively equivalent, for all types of errors in Tab.\,\ref{tab:error}, with a slight advantage of the $\sigma_1$ with respect to the $\sigma_2$ control. Moreover, fixing the peak, type (2) error in Tab.\,(\ref{tab:error}), is preferable over the others, improving the average probability by up to a factor of three. Overall, we confirm here that the $\sigma_1$ control is advantageous when we deal with a distribution of gaps.



\subsection{Modifying the pulse function}

\begin{figure}[!htb]
    \centering
    \includegraphics[width=1.\linewidth]{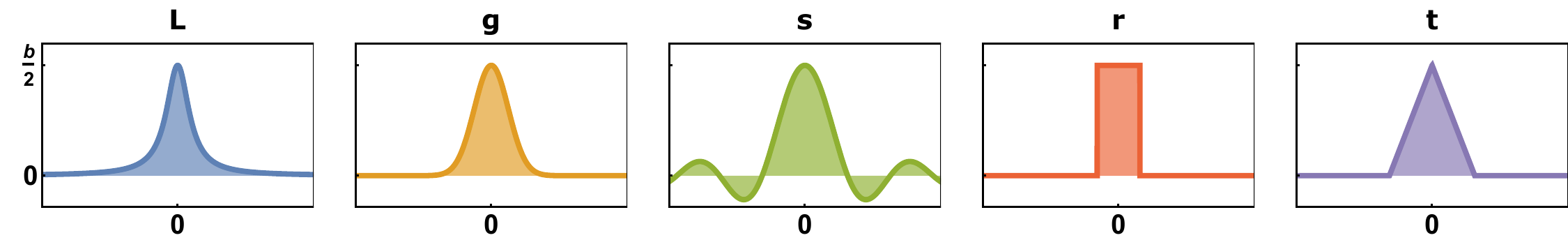}
    \caption{Graphs of $f_\text{D}(t;b)$ from Tab. \ref{tab:pref-examples} as a function of time. At $t = 0$, $f_\text{D}(t;b) = b/2$ and the area under the graph, shaded region, is equal to $\pi/2$. The area-peak conditions, Eq. (\ref{eq:rea-peak}), specify the functional dependence within a certain family of functions, e.g, for the rectangular pulse, the width and height as a function of $b$.
    }
    \label{fig:9}
\end{figure}

In the generalized Hamiltonian of Eq. (\ref{eq:GLZS}), we propose an additional modification, changing the shape of the counterdiabatic pulse $f_\text{CD}$, see Eq. (\ref{eq:fCD}), to a more general form $H_1 = f_\text{D}\sigma_\varphi$. Modeled after the original Lorentzian shape, $f_\text{D}^{\,\texttt{L}} = \frac{b/2}{b^2t^2+1}$, we use various functions as shown in Tab. \ref{tab:pref-examples} and plotted in Fig. \ref{fig:9}, which are otherwise chosen arbitrarily. As is the case for $\texttt{L}$, the exact functional form is fixed by imposing the conditions
\begin{equation}
    \int_\mathbb{R}\text{d}t\ f_\text{D}(t;b) = \frac{\pi}{2},\quad f(0;b) = \frac{b}{2}
    \label{eq:rea-peak}
\end{equation}

\begin{table}[!h]
\def\arraystretch{1.}
    \centering
    \begin{tabular}{r|l}
    \toprule
     \texttt{\#}& $f_\text{D}(t;b)$\\
    \midrule
         \texttt{L} &  $\frac{b/2}{b^2t^2+1}$\\
         \texttt{g} &  $\frac{b}{2}\text{e}^{-b^2t^2/\pi}$\\
         \texttt{s} &  $\frac{1}{2}\frac{\sin bt}{t}$\\
         \texttt{r} &  $\frac{b}{2}\mathbf{1}(\abs{t}<\tfrac{\pi}{2b})$\\
         \texttt{t} &  $\frac{1}{2}\mathbf{1}(\abs{t}<\tfrac{\pi}{b})(b-\frac{\abs{t}}{\pi}b^2)$\\
    \bottomrule
    \end{tabular}
    \caption{$b>0$. Example of general pulse functions modeled after the Lorentzian \texttt{L}.
    }
    \label{tab:pref-examples}
\end{table}

Moreover, the pulses $f_\text{D}$ are designed such that they preserve the limit $b\rightarrow \infty$, see the red curve in Fig. \ref{fig:1}. In this way, $f_\text{D}$ controls the $a\rightarrow 0$ limit, see the blue curve in Fig. \ref{fig:1}, and $\varphi$ controls the $b\rightarrow \infty$ limit independently of one another, see \ref{ap:angle}.

\begin{figure}[!ht]
    \centering
    \includegraphics[width=1.\linewidth]{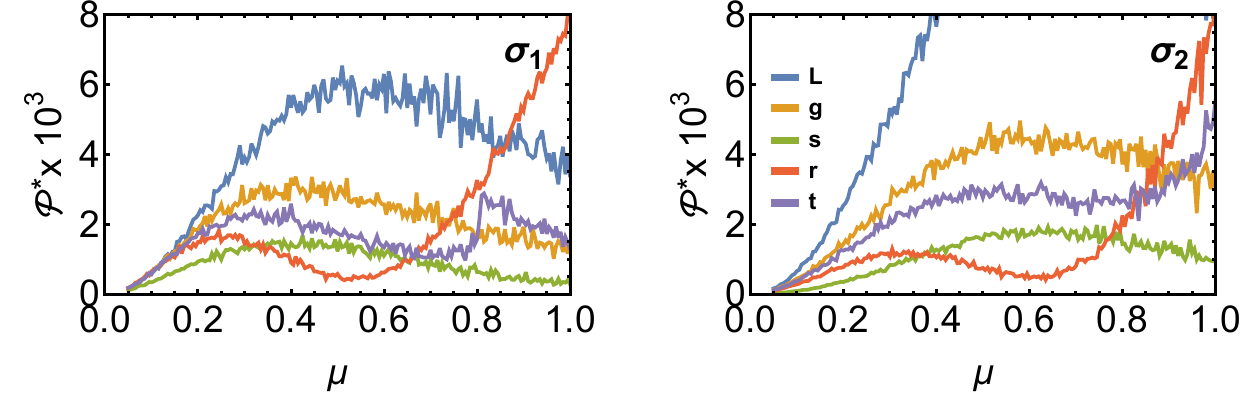}
    \caption{$\mathcal{P}^\ast$ as a function of $\mu$, with $a\sim \mathcal{N}(\mu,\mu^2/25)$. For small $\mu$, \texttt{L} is the worst among our examples from Tab. \ref{tab:pref-examples}. $\varphi = 0,\ \sigma_1$ (left panel) and $\varphi = \pi/2,\ \sigma_2$ (right panel). Comparing the $\sigma_1$ and $\sigma_2$ cases, for almost all values of $\mu$, $\mathcal{P}^\ast[\sigma_1]<\mathcal{P}^\ast[\sigma_2]$, independently of the pulse form $f_\text{D}$.}
    \label{fig:10}
\end{figure}

In Fig. \ref{fig:10}, we have plotted $\mathcal{P}^\ast$ as a function of $\mu$, $a\sim \mathcal{N}(\mu,\mu^2/25)$ for $\sigma_1$ (left panel) and $\sigma_2$ (right panel). For small $\mu$, the original Lorentzian pulse is the worst among our list of functions from Tab. \ref{tab:pref-examples}. The general correction $H_1$ suffers from non-zero transition probabilities, similar to what is seen in the left panel of Fig. \ref{fig:2}. For large $\mu$,  the corresponding Characteristic Curve, see our definition in Eq. (\ref{eq:CC}), becomes irregular and it is typically multivalued. Thus, extra care must be taken in extending the results to larger $\mu \gtrsim 1$. The numerical analysis again shows an advantage of $\sigma_1$ with respect to $\sigma_2$, for almost all values $\mu\in(0,2)$ and all functions $f_\text{D}$ in Tab. \ref{tab:pref-examples}. Overall, the choice \texttt{s} performs best, and the original \texttt{L} performs worst in this parameter range. \mfaC{From the experimental point of view, the Gaussian pulse \texttt{g} might be the best compromise between performance and reliability.} Further engineering and optimization of the pulse $f_\text{D}$ is the subject of future work.

\subsection{Nonlinear sweeps}

In Eq. (\ref{eq:LZST}), appropriately designed nonlinear sweeps \cite{nlRC, Lidar2010, nl-tan, bason1, Polkovnikov2016, wimberger-19, polk-geometry, Dengis2025, Dengis2025-2, Romanato2025} are advantageous to the linear ramp, allowing for a significant speed-up of the adiabatic evolution, i.e., for small $T$. \mfaC{Nonlinear models are rarely exactly solvable and can only be treated analytically in the adiabatic or the perturbative limit \cite{vitanov-nls,LZ-nls}.}
One of the disadvantages is that their dependence on the parameters is typically highly nonlinear, which is a drawback in our case since the gap $a$ is not fixed.

\begin{figure}[!tbh]
    \centering
    \includegraphics[width=1.\linewidth]{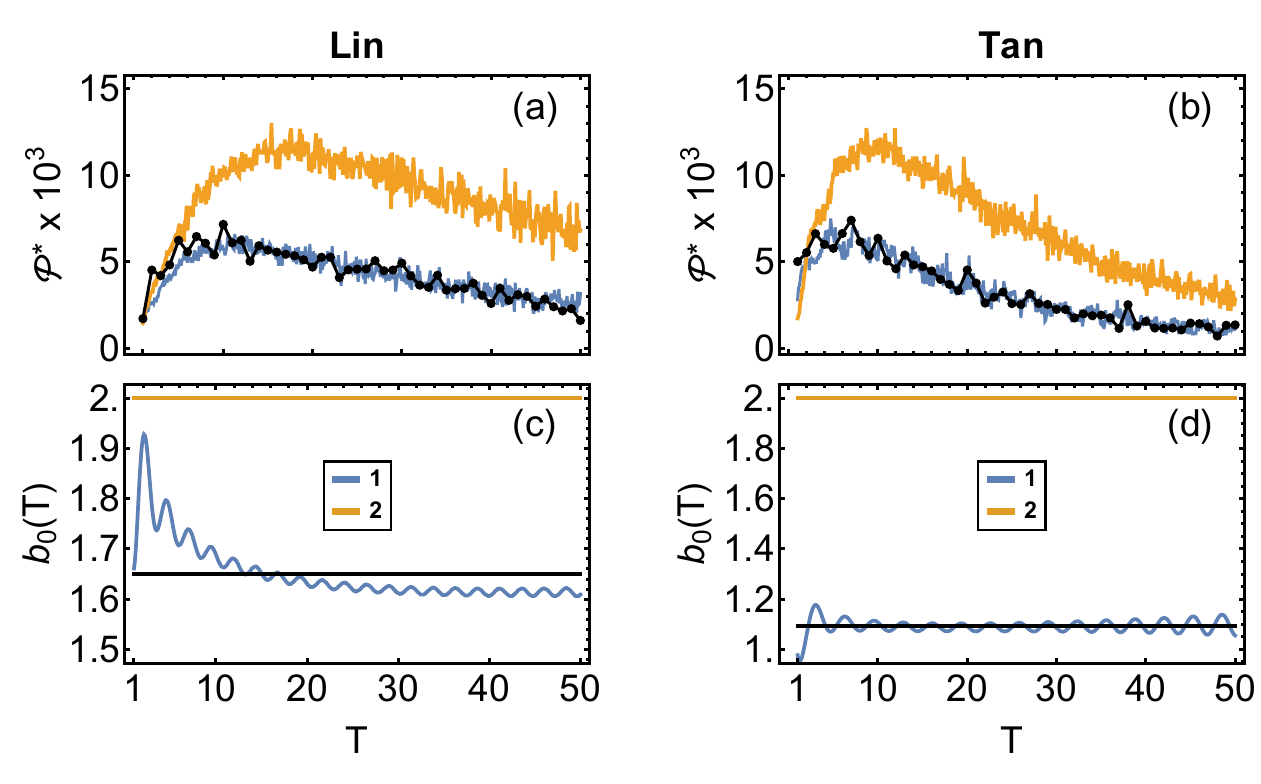}
    \caption{(Top row) $\mathcal{P}^\ast$ as a function of protocol time $T$, for a distribution of gaps $a\sim \mathcal{N}(\mu,\mu^2/25)$ with $\mu = 0.5$ and $\lambda_0 = 10$, Eq. (\ref{eq:LZST2}). We applied the sweep functions stated in Eq. (\ref{eq:NLS}), right panel (a): linear sweep, left panel (b): tangent sweep. (Bottom row) $b_0$ as a function of $T$. For $\sigma_1$ we observe oscillations whereas for $\sigma_2$ there is no $T$ dependence. The colors correspond to the controls $\sigma_1$ (blue lines) and $\sigma_2$ (orange lines). Black points/line in panels (a) and (b) correspond to the average value $b_0(T)$ indicated by the black lines in (c) and (d).
    }
    \label{fig:11}
\end{figure}

Figure \ref{fig:11} shows a comparison between a linear (\texttt{Lin}) and a tangent sweep (\texttt{Tan}) with
\begin{equation}
    \lambda_\texttt{Lin}(s) = \lambda_0(s-\tfrac{1}{2}),\quad \lambda_\texttt{Tan}(s;a) = \frac{\lambda_0}{2}a \tan(\arctan(\tfrac{1}{a})(2s-1)),
    \label{eq:NLS}
\end{equation}
$a$ being the gap parameter in Eq. (\ref{eq:LZST}). \mfaC{The exact form of $\lambda_\texttt{Tan}$ is fixed by matching its values with $\lambda_\texttt{Lin}$ at the boundaries of the interval $s\in[0,1]$.} As is the case for the correction term $H_1$, the sweep is not part of the system that fluctuates, and therefore the parameter $a$ in $\lambda(s;a)$ should be kept constant. Eq. (\ref{eq:LZST}) now takes the form
\begin{equation}
    i\partial_s \ket{\psi} = \bigg[T(-\lambda(s;c) \sigma_3 +a\sigma_1) + \partial_s\lambda(s;b)\frac{b/2}{b^2\lambda(s;b)^2+1}\sigma_\varphi\bigg]\ket{\psi},\ s\in[0,1],
    \label{eq:LZST2}
\end{equation}
introducing a new parameter $c$ that is independent of the gap. Similar to Eq. (\ref{eq:bstar}), given $a\sim\mathcal{N}(\mu,\sigma^2)$, $c(\mu,\sigma)$ can be optimized for the specific values $\mu,\sigma$. For simplicity, we will set $c = \mu$, ignoring a potential $\sigma$ dependence. As it was done previously, for the coupling $b$ we will use $b_0$ with the caveat that for $\varphi\neq\pi/2$, $b_0$ now also depends on $T$, see panels (c) and (d) in Fig. \ref{fig:11}. Our numerical analysis indicates that we can safely ignore this, in a wide range of $T$. Since the evolution for $\sigma_1$ is not transitionless in the center, see the left panel in Fig. \ref{fig:3}, the dependence of $b_0$ on $T$, see the blue curves in panels (c) and (d) in Fig. \ref{fig:11}, shows oscillating behavior that is slightly better for the $\texttt{Tan}$ sweep. For small $T$, the two sweeps perform similarly with the $\texttt{Tan}$ sweep overtaking \texttt{Lin} for larger $T$. As in the previous sections, the advantage of $\sigma_1$ over $\sigma_2$ is small but systematic.  

We conclude that for a nonlinear sweep using the $\sigma_1$ control is again giving better results for a LZ problem with a random distribution of gaps, but little or no improvement is gained with respect to a linear sweep.

\section{Conclusions}
\label{sec:concl}

We have investigated in detail the effect of parameter fluctuations, modeled by a random distribution of gaps, in the control of a LZ transition. Our goal was to find the best control pulse, inspired by counterdiabatic control \cite{berry-trans, demirplak1, Guery}, that on average best suppresses non-adiabatic transitions. The upshot is that an additional control field based on real matrix elements, i.e. by $\sigma_1$, is better than the standard counterdiabatic $\sigma_2$ result. Since real matrix elements are already present in the original Hamiltonian, this has the advantage that no new matrix elements have to be introduced and controlled by the experimenter, similar to the motivation for the effective counterdiabatic protocols by Petiziol et al., see Refs. \cite{wimberger-18, wimberger-24}. In combination with an optimization of the pulse form, the speed of the LZ protocol can be significantly accelerated with respect to the case without additional controls. 

Our results, mostly proven for a Normal distribution of couplings, generalize to unimodal distributions of parameter uncertainties. This demonstrates that, over a broad range of system and fluctuation parameters, we can indeed control also systems with random parameters. Such control is important for any experiments subject to a priori unknown couplings, and may provide new pathways for quantum technologies that exploit adiabatic transfers to be robust against variations in the gap-parameter \cite{Bateman2007}. It may also allow for control of the spin-collision dynamics of two thermal atoms in optical tweezers \cite{exper2}. Similar experiments were performed with two qubits controlled by nuclear magnetic resonance \cite{Suresh2023}, for which counterdiabatic control was applied. Since it is likely that all such implementations suffer to some degree from random parameter fluctuations, this gives us hope for practical applications of the methods developed here.

\mfaC{The standard LZ model with fluctuating parameters has been studied extensively \cite{LZD-kayanuma,kayanuma-offdiagonal,Ao1989,Wubs2006,Kayanuma-diss,Huang2018,Pokrovsky-fastNoise}.
When the time correlations are much faster than the time scales of the system, typically with $\delta$-correlated white noise, the average behavior is captured by Lindbladian dynamics whereas for finite-time correlations, there is need for a non-Markovian description \cite{Kiely2021}. Our case corresponds to the static limit in which the parameter fluctuations are much slower that the typical time scale. Future work will focus on the the interplay of noise with finite-time correlations and the control with a counterdiabatic field. In doing so, the formalism of transitionless driving must be extended to open quantum systems \cite{CDopen-Vacanti,CDopen-Query,CDopen-Alipour,CDopne-Santos} with many questions still being open.}

\section*{Acknowledgements}
We thank Lucia Deimel, Francesco Petiziol, and Robert Weiß for comments on our work and the manuscript.
MFA and SW acknowledge funding by Q-DYNAMO (EU HORIZON-MSCA-2022-SE-01) with project No. 101131418, and SW by the National Recovery and Resilience Plan, through PRIN 2022 project "Quantum Atomic Mixtures: Droplets, Topological Structures, and Vortices", project No. 20227JNCWW, CUP D53D23002700006, and through Mission 4 Component 2 Investment 1.3, Call for tender No. 341 of 15/3/2022 of Italian MUR funded by NextGenerationEU, with project No. PE0000023, Concession Decree No. 1564 of 11/10/2022 adopted by MUR, CUP D93C22000940001, Project title "National Quantum Science and Technology Institute" (NQSTI). MFA also acknowledges support from Quantum Technologies Aotearoa, a research programme of Te Whai Ao—the Dodd Walls Centre, funded by the New Zealand Ministry of Business Innovation and Employment through International Science Partnerships, Contract No. UOO2347.

\appendix
\section{}
\label{ap:LZS}

We start with some comments on the notation used in the following: For the LZ system, Eq. (\ref{eq:LZS}), we denote the time evolution matrix by $\mathcal{U}_0$, the matrix elements of $\mathcal{U}_0$ with $\mathcal{A}_0,\mathcal{B}_0$ as in Eq. (\ref{eq:LZU}) and the corresponding transition probability $\mathcal{P}_0 = \abs{\mathcal{A}_0}^2$. For the generalized Hamiltonian, Eq. (\ref{eq:GLZS}), we use analogous notation, preserving the subscript $0$ for the case $b = 0$ and $\infty$ for the case $b\rightarrow\infty$. The dependence on the parameter $\varphi$ is usually omitted. Specifically, for the time evolution in intervals of the form
\[
    [-t_f,0],\quad [0,t_f],\quad t_f>0
\]
we use the subscripts $\pm$.
\subsection{Landau Zener}
For Eq.\,(\ref{eq:LZS}), the formulas for $\mathcal{A}_0,\mathcal{B}_0$ can be found in \cite{vitanov-finite,nori2}. For convenience, we collect all the relevant equations
\begin{align}
    \mathcal{A}_0(t_f,t_i;a)
        &= \frac{\Gamma(1-\nu)}{\sqrt{2\pi}}\bigg[D_{\nu}(z_f)D_{\nu-1}(-z_i)+D_\nu(-z_f)D_{\nu-1}(z_i)\bigg] \label{eq:LZSA}\\
    \mathcal{B}_0(t_f,t_i;a)
        &= \frac{\Gamma(1-\nu)}{a\sqrt{\pi}}e^{i\frac{\pi}{4}}\bigg[-D_{\nu}(z_f)D_{\nu}(-z_i)+D_\nu(-z_f)D_{\nu}(z_i)\bigg]\label{eq:LZSB}
\end{align}
\begin{equation}
    \nu\equiv i\frac{a^2}{2},\quad z_i \equiv \sqrt{2}e^{-i\frac{\pi}{4}}t_i,\quad z_f \equiv \sqrt{2}e^{-i\frac{\pi}{4}}t_f \label{eq:LZSNZ}
\end{equation}
with $D_\nu(z)$ being the Parabolic Cylinder Functions \cite{whittaker}. From their asymptotic expansions for large $\abs{t_i},\abs{t_f}$ \cite{asymp,whittaker,vitanov-book}, the leading terms are
\begin{equation}
    \mathcal{A}_0 \sim \text{e}^{-\pi\frac{a^2}{2}+i(\varphi_f-\varphi_i)},\quad
    \mathcal{B}_0 \sim -\sqrt{1-\text{e}^{-\pi a^2}}e^{i\theta}
    \label{eq:LZAB0}
\end{equation}
\begin{equation}
    \theta(t_f,t_i)\equiv \varphi_i+\varphi_f + \frac{\pi}{4} + \arg\Gamma(1-\nu)
\end{equation}
\begin{equation}
    \varphi_i \equiv \tfrac{1}{2}(t_i^2+\tfrac{1}{2}a^2\ln 2t_i^2),\quad \varphi_f \equiv \tfrac{1}{2}(t_f^2+\tfrac{1}{2}a^2\ln 2t_f^2)
\end{equation}
Setting $t_f = -t_i$ in Eq. (\ref{eq:LZAB0}), we recover the Landau-Zener formula
\begin{equation}
    \mathcal{P}_0(a) = |\mathcal{A}_0(a)|^2 = \text{e}^{-\pi a^2}
    \label{eq:PLZ0}
\end{equation}

\subsection{Symmetries}

Eq. (\ref{eq:LZS}) exhibits the following properties that also generalize to $b\neq 0$, Eq. (\ref{eq:GLZS}), $i\partial_t\ket{\psi} = H_\text{GLZ}\ket{\psi}$
\begin{equation}
    i\partial_t \ket{\psi} = \bigg[-t\sigma_3+a\sigma_1 + \frac{b/2}{b^2t^2+1}\sigma_\varphi\bigg]\ket{\psi}
    \label{eq:GLZSEq}
\end{equation}
with $\sigma_\varphi\equiv\cos\varphi \sigma_1+\sin\varphi\sigma_2$.
\begin{enumerate}
\item 
The time evolution matrices $\mathcal{U}(t,0)$ and $\mathcal{U}(0,-t),t>0$ are related.
Let
\[
    \tau \equiv -t,\quad \ket{\varphi(\tau)} \equiv\sigma_3\ket{\psi(-\tau)}
\]
Substituting $\tau$ in Eq. (\ref{eq:GLZSEq}) we get 
\begin{align*}
    i\partial_\tau \ket{\psi(-\tau)}
        &= \bigg(-\tau\sigma_3-a\sigma_1-\frac{b/2}{b^2\tau^2+1}\sigma_\varphi\bigg)\ket{\psi(-\tau)}
\end{align*}
Multiplying both sides by $\sigma_3$ and inserting the identity $\sigma_3^2 = \mathbf{1}$
\begin{align*}
        i\partial_\tau (\sigma_3\ket{\psi(-\tau)})
        &= \sigma_3\bigg(-\tau\sigma_3-a\sigma_1-\frac{b/2}{b^2\tau^2+1}\sigma_\varphi\bigg)\sigma_3^2\ket{\psi(-\tau)}
\end{align*}
Substituting $\ket{\varphi(\tau)}$
\begin{align*}
    i\partial_\tau \ket{\varphi(\tau)}
        &= \sigma_3\bigg(-\tau\sigma_3-a\sigma_1-\frac{b/2}{b^2\tau^2+1}\sigma_\varphi\bigg)\sigma_3 \ket{\varphi(\tau)}
\end{align*}
Finally, using the Pauli matrices relations
\[
    \quad \sigma_3\sigma_\varphi\sigma_3 = -\sigma_\varphi
\]
and by relabeling $\tau\rightarrow t$ we get 
\begin{equation}
    i\partial_t \ket{\varphi(t)}= \bigg(-t \sigma_3+a\sigma_1+\frac{b/2}{b^2t^2+1}\sigma_\varphi\bigg)\ket{\varphi(t)}
\end{equation}
i.e, $i\partial_t\ket{\varphi} = H_\text{GLZ}\ket{\varphi}$. The solution to the last equation, in the interval $[0,t]$ is
\[
    \ket{\varphi(t)} = \mathcal{U}(t,0)\ket{\varphi(0)}
\]
with $\mathcal{U}$ the corresponding time evolution matrix. Expressing the result in term of $\ket{\psi(t)}$
\[
            \ket{\psi(0)} = \sigma_3 \mathcal{U}^\dagger(t,0) \sigma_3 \ket{\psi(-t)}
\]
Comparing with $\ket{\psi(0)} = \mathcal{U}(0,-t)\ket{\psi(-t)}$ we get
\[
    \sigma_3 \mathcal{U}^\dagger(t,0) \sigma_3 = \mathcal{U}(0,-t)
\]
Expanding the matrices and using the shorthand notation
\[
    \begin{pmatrix} 1 &0\\0&-1\end{pmatrix}
    \begin{pmatrix} \mathcal{A}_{+}^\ast &-\mathcal{B}_{+} \\ \mathcal{B}_{+}^\ast & \mathcal{A}_{+} \end{pmatrix}
    \begin{pmatrix} 1 &0\\0&-1\end{pmatrix}
    \stackrel{!}{=}\begin{pmatrix} \mathcal{A}^\ast_{-} &\mathcal{B}_{-} \\ -\mathcal{B}^\ast_{-} & \mathcal{A}_{-} \end{pmatrix}
\]
we get
\begin{equation}
    \mathcal{A}_{-} = \mathcal{A}^\ast_{+},\quad \mathcal{B}_{-} = \mathcal{B}_{+}
    \label{eq:sym1}
\end{equation}
\item 
The LZ formula, Eq. (\ref{eq:PLZ0}) is even in $a$: $\mathcal{P}_0(-a) = \mathcal{P}_0(a)$ which generalizes for Eq. (\ref{eq:GLZSEq}) to
\begin{equation}
    \mathcal{P}(a,b) = \mathcal{P}(-a,-b)
    \label{eq:sym2}
\end{equation}
For the proof, using $\sigma_3\sigma_\varphi\sigma_3 = -\sigma_\varphi$ 
\[
    H_\text{GLZ}(t;a,b) = -t\sigma_3+a\sigma_1 +\frac{b/2}{b^2t^2+1}\sigma_\varphi= \sigma_3 H_\text{GLZ}(t;-a,-b)\sigma_3
\]
from which follows
\[
    \mathcal{U}(t_f,t_i;a,b) = \sigma_3 \mathcal{U}(t_f,t_i;-a,-b)\sigma_3
\]
and subsequently, for the diagonal elements $\mathcal{A} \equiv \mathcal{U}_{11}$
\[
    \mathcal{A}(t_f,t_i;a,b) = \mathcal{A}(t_f,t_i;-a,-b)
\]
which proves Eq. (\ref{eq:sym2}) by trivially taking the limits $t_f\rightarrow + \infty, t_i\rightarrow -\infty$ and $\mathcal{P} = \abs{\mathcal{A}}^2$.
\item
In Eq. (\ref{eq:LZAB0}), letting $t_f = -t_i$ leads to $\mathcal{A}_0\in\mathbb{R}$.
This is a consequence of (i) and the property 
$\mathcal{U}(t_f,t_i) = \mathcal{U}(t_f,t_0)\mathcal{U}(t_0,t_i)$. 
For $t_f = -t_i = t>0$ and $t_0 = 0$
\[
    \mathcal{U}(t,-t) = \mathcal{U}(t,0) \mathcal{U}(0,-t)
\]
Expanding the matrices we obtain
\begin{align*}
    \mathcal{A} &= \mathcal{A_{+}}\mathcal{A}_{-}-\mathcal{B}_{+}\mathcal{B}_{-}^\ast \\
    \mathcal{B} &= \mathcal{A_{+}}\mathcal{B}_{-}+\mathcal{A}_{-}^\ast\mathcal{B}_{+}
\end{align*}
In combination with Eq. (\ref{eq:sym1})
\begin{equation}
    \mathcal{A} = \abs{\mathcal{A_{+}}}^2-\abs{\mathcal{B}_{+}}^2 = 2\abs{\mathcal{A_{+}}}^2-1 \in\mathbb{R} \label{eq:sym3}
\end{equation}
\begin{equation}
    \mathcal{B} = \mathcal{A_{+}}\mathcal{B}_{+}+\mathcal{A}_{-}^\ast\mathcal{B}_{+}
        = 2\mathcal{A_{+}}\mathcal{B}_{+}
    \label{eq:sym4}
\end{equation}
\end{enumerate}


\section{}
\label{ap:red}
\subsection{Landau Zener with Dirac-$\delta$}
\label{ap:red_LZD}

Let $\boldsymbol{n} = (n_1,n_2,n_3)\in\mathbb{R}^3$ and
\begin{equation}
    i\partial_t\ket{\psi} = \bigg[-t\sigma_3+a\sigma_1+\delta(t)(\boldsymbol{n}\cdot \boldsymbol{\sigma})\bigg]\ket{\psi},\quad a\neq 0
    \label{eq:LZdelta}
\end{equation}
Integrating Eq. (\ref{eq:LZdelta}) in the symmetric interval $[-t_f,t_f]$, with $t_f>0$, the time evolution matrix becomes
\begin{equation}
    \mathcal{U}_\infty(t_f,-t_f;a,\boldsymbol{n}) = \mathcal{U}_0(t_f,0;a)\mathcal{K}(\boldsymbol{n})\mathcal{U}_0(0,-t_f;a)
    \label{eq:ap-redU}
\end{equation}
with the factor \ref{ap:DC}
\begin{equation}
    \mathcal{K}(\boldsymbol{n})
        \equiv \text{e}^{-i\ (\boldsymbol{n}\cdot\boldsymbol{\sigma})}
        = \cos\abs{\boldsymbol{n}} -i\sin \abs{\boldsymbol{n}}(\hat{\boldsymbol{n}}\cdot\boldsymbol{\sigma})
        \label{eq:ap-kappa}
\end{equation}
arising due to the Dirac $\delta(t)$ and $\mathcal{U}_0$ as in Eq. (\ref{eq:LZU}).
Expanding the matrices in Eq. (\ref{eq:ap-redU})
\begin{equation}
    \mathcal{A}_\infty
        = C_0 \cos\abs{\boldsymbol{n}}
            -i \frac{n_3}{\abs{\boldsymbol{n}}}\sin\abs{\boldsymbol{n}}C_3
            +\sin\abs{\boldsymbol{n}}\left(\frac{n_1}{\abs{\boldsymbol{n}}}C_1+\frac{n_2}{\abs{\boldsymbol{n}}}C_2\right)
    \label{eq:ap-redA}
\end{equation}
with $\mathcal{A}_\infty$ being the diagonal elements of $\mathcal{U}_\infty$ and $C_i(a),i=0,1,2,3$ given by
\begin{align}
    C_0 &= (\mathcal{U}_{0,+}\mathcal{U}_{0,-})_{11}
            = \mathcal{A}_0
            \label{eq:C0}\\
    C_1 &= -i(\mathcal{A}_{0,-}\mathcal{B}_{0,+}-\mathcal{A}_{0,+}\mathcal{B}_{0,-}^\ast)
            =2\text{Im}(\mathcal{A}_{0,-}\mathcal{B}_{0,-})
            \label{eq:C1}\\
    C_2 &= \mathcal{A}_{0,-}\mathcal{B}_{0,+}+\mathcal{A}_{0,+}\mathcal{B}_{0,-}^\ast
            =2\text{Re}(\mathcal{A}_{0,-}\mathcal{B}_{0,-})
            \label{eq:C2}\\
    C_3 &= \mathcal{A}_{0,+}\mathcal{A}_{0,-}+\mathcal{B}_{0,+}\mathcal{B}_{0,-}^\ast
            =\abs{\mathcal{A}_{0,+}}^2+\abs{\mathcal{B}_{0,+}}^2 =1
            \label{eq:C3}
\end{align}
with the second equalities following from Eq. (\ref{eq:sym1}).
$\mathcal{A}_0$, $\mathcal{A}_{0,-}$ and $\mathcal{B}_{0,-}$ are given by Eqs. \eqref{eq:LZSA} and \eqref{eq:LZSB}
\begin{align}
    \mathcal{A}_0 &\equiv \mathcal{A}_0(t_f,-t_f;a) = \frac{\Gamma(1-\nu)}{\sqrt{2\pi}}\bigg(D_\nu(z_f)D_{\nu-1}(z_f)+D_\nu(-z_f)D_{\nu-1}(-z_f)\bigg)\\
    \mathcal{A}_{0,-} &\equiv \mathcal{A}_0(0,-t_f;a) =  \frac{\Gamma(1-\nu)}{\sqrt{2\pi}}D_\nu(0)\bigg(D_{\nu-1}(z_f)+D_{\nu-1}(-z_f)\bigg) \label{eq:A0-}\\
    \mathcal{B}_{0,-} &\equiv \mathcal{B}_0(0,-t_f;a) =  \frac{\Gamma(1-\nu)}{a\sqrt{\pi}}e^{i\frac{\pi}{4}}D_\nu(0)\bigg(-D_{\nu}(z_f)+D_{\nu}(-z_f)\bigg)\label{eq:B0-}
\end{align}
In the limit $t_f\rightarrow +\infty$, keeping leading order terms in $t_f$ \cite{asymp,whittaker,vitanov-book} 
\begin{align}
    \mathcal{A}_0 &\sim \text{e}^{-\pi a^2/2}\\
    \mathcal{A}_{0,-} &\sim D_\nu(0)\text{e}^{-\pi a^2/8-i\varphi_f}\\
    \mathcal{B}_{0,-} &\sim \frac{\Gamma(1-\nu)}{a\sqrt{\pi}}D_\nu(0)(\text{e}^{-3\pi a^2/8}-\text{e}^{\pi a^2/8})e^{i(\varphi_f+\frac{\pi}{4})} 
\end{align}
In Eqs. (\ref{eq:C2}) and (\ref{eq:C3}), only the product $\mathcal{A}_{0,-}\mathcal{B}_{0,-}$ appears in which $t_f$ drops out in the leading order term
\begin{equation}
    \mathcal{A}_{0,-}\mathcal{B}_{0,-}
        \sim\frac{\Gamma(1-\nu)}{a\sqrt{\pi}}D_\nu^2(0)(\text{e}^{-\pi a^2/2}-1)e^{i\frac{\pi}{4}}
    \label{eq:AB0-1}
\end{equation}
Substituting $D_\nu(0) =2^{\frac{\nu}{2}}\sqrt{\pi}/\Gamma\left(\frac{1-\nu}{2}\right)$, \cite[Chapter 19]{abramowitz} and after further algebraic manipulations we get
\begin{align}
    \mathcal{A}_{0,-}\mathcal{B}_{0,-}
        &\sim \frac{2^\nu\sqrt{\pi}}{a}\frac{\Gamma(1-\nu)}{\Gamma^2\left(\tfrac{1-\nu}{2}\right)}
                (\text{e}^{-\pi a^2/2}-1)e^{i\frac{\pi}{4}}\nonumber \\
        &= \frac{\sqrt{\pi}}{a} (\text{e}^{-\pi a^2/2}-1)\frac{|\Gamma(1-\nu)|}{|\Gamma(\frac{1-\nu}{2})|^2}
                e^{i\omega(a)}
    \label{eq:AB0-2}
\end{align}
with 
\begin{equation}
    \omega(a) \equiv \frac{\pi}{4} + \frac{a^2}{2}\ln 2 + \arg\Gamma(1-\nu) - 2\arg\Gamma\left(\frac{1-\nu}{2}\right)
    \label{eq:omega}
\end{equation}
Let $x \equiv \pi \frac{a^2}{2}$. Using the formula $\sinh(x)=2\sinh(\tfrac{x}{2})\cosh(\tfrac{x}{2})$
and properties from Eqs. (\ref{eq:gamma5}) and (\ref{eq:gamma6}),
the ratio of $\Gamma$ functions in \eqref{eq:AB0-2} becomes
\begin{align*}
    \frac{|\Gamma(1-\nu)|}{|\Gamma(\tfrac{1-\nu}{2})|^2}
        & = \frac{|a|}{\sqrt{2\pi}}\frac{\cosh \tfrac{x}{2}}{\sqrt{\sinh x}}
        = \frac{|a|}{2\sqrt{\pi}}\frac{1}{\sqrt{\tanh \tfrac{x}{2}}}
\end{align*}
from which 
\begin{equation}
    \mathcal{A}_{0,-}\mathcal{B}_{0,-}
        \sim \frac{\abs{a}}{2a}\frac{e^{-x}-1}{\sqrt{\tanh\tfrac{x}{2}}}e^{i\omega(a)}
        = -\frac{1}{2}\text{sgn}(a)\sqrt{1-e^{-\pi a^2}}e^{i\omega(a)}
    \label{eq:AB0-}
\end{equation}
with $\text{sgn}(a)\equiv\frac{\abs{a}}{a}$.
Eq. \eqref{eq:omega} can be written in an equivalent but more familiar form.
From the Legendre duplication formula, Eq. (\ref{eq:gamma3}) with $z =(1-\nu)/2$
\[
    \frac{\Gamma(1-\nu)}{\Gamma(\frac{1-\nu}{2})} =\frac{1}{\sqrt{\pi}} 2^{-\nu}\Gamma(\tfrac{2-\nu}{2})
\]
from which follows
\begin{align*}
    \arg \frac{\Gamma(1-\nu)}{\Gamma(\tfrac{1-\nu}{2})}
        &= \arg(2^{-\nu}\Gamma(\tfrac{2-\nu}{2})/\sqrt{\pi})
            = \arg(2^{-\nu}\Gamma(\tfrac{2-\nu}{2}))\\
        &=\arg 2^{-i a^2/2} + \arg \Gamma(\tfrac{2-\nu}{2})\\
        &= - \frac{a^2}{2}\ln 2 + \arg \Gamma(\tfrac{2-\nu}{2})
\end{align*}
and 
\[
    \arg\Gamma(1-\nu)-\arg \Gamma(\tfrac{1-\nu}{2})
        = -\frac{a^2}{2}\ln2 + \arg \Gamma(\tfrac{2-\nu}{2})
\]
Substituting the last equation in Eq. \eqref{eq:omega} we get $\omega(a) = \frac{\pi}{2}-\chi(a)$ with
\begin{align}
    \chi(a) &\equiv \frac{\pi}{4} +\arg\Gamma\left(\frac{1-\nu(a)}{2}\right)-\arg\Gamma\left(\frac{2-\nu(a)}{2}\right)\nonumber \\
        \label{eq:ap-chi}
\end{align}
Combining all the above with Eq. \eqref{eq:ap-redA} we have 
\begin{multline}
    \mathcal{A}_\infty(a) = \text{e}^{-\pi a^2/2}\cos|\mathbf{n}|
        - i \frac{n_3}{|\mathbf{n}|}\sin|\mathbf{n}|\\
        -\text{sgn}(a)\sqrt{1-\text{e}^{-\pi a^2}}\sin|\mathbf{n}|\bigg[\frac{n_1}{|\mathbf{n}|} \cos\chi(a)+\frac{n_2}{|\mathbf{n}|}\sin\chi(a)\bigg]
    \label{eq:redProb2}
\end{multline}
For  $n_3 = 0$, let $\mathbf{n} = \abs{\mathbf{n}}(\cos\varphi,\sin\varphi,0)$, Eq. (\ref{eq:redProb2}) reduces to 
\begin{equation}
    \mathcal{A}_\infty(a) = \text{e}^{-\pi a^2/2}\cos\abs{\boldsymbol{n}}
        -\text{sgn}(a)\sqrt{1-\text{e}^{-\pi a^2}}\sin\abs{\boldsymbol{n}} \cos(\chi(a)-\varphi)
    \label{eq:redProb3}
\end{equation}
and for $\abs{\mathbf{n}} = \frac{\pi}{2}$
\begin{equation}
    \mathcal{A}_\infty(a) = -\text{sgn}(a)\sqrt{1-\text{e}^{-\pi a^2}}\cos(\chi(a)-\varphi)
\end{equation}

\begin{equation}
    \mathcal{P}_\infty(a) = (1-\mathcal{P}_0(a))\cos^2(\chi(a)-\varphi)
\end{equation}
with $\mathcal{P}_0 = \text{e}^{-\pi a^2}$ and $\chi(a)$ given by Eq. (\ref{eq:ap-chi}).

\section{Other calculations}
\subsection{Discontinuity Matrix}
\label{ap:DC}
Derivation of Eq. (\ref{eq:ap-kappa})

For $\boldsymbol{a,b,n}\in\mathbb{R}^3$, the Pauli Matrices satisfy
\[
    (\boldsymbol{a}\cdot\boldsymbol{\sigma})(\boldsymbol{b}\cdot\boldsymbol{\sigma}) = (\boldsymbol{a}\cdot \boldsymbol{b})\mathbf{1}
        + i (\boldsymbol{a}\times\boldsymbol{b})\cdot \boldsymbol{\sigma}
\]
from which 
\[
    (\boldsymbol{n}\cdot \boldsymbol{\sigma})^{2k} = \abs{\boldsymbol{n}}^{2k}\mathbf{1},\quad
        (\boldsymbol{n}\cdot \boldsymbol{\sigma})^{2k+1} = \abs{\boldsymbol{n}}^{2k+1}(\hat{\boldsymbol{n}}\cdot \boldsymbol{\sigma})
\]
Splitting the series of the exponential $\text{exp}(-i\boldsymbol{n}\cdot \boldsymbol{\sigma})$ in even/odd powers we get
\[
    \text{exp}(-i\boldsymbol{n}\cdot \boldsymbol{\sigma}) = \cos\abs{\boldsymbol{n}}\mathbf{1}-i(\hat{\boldsymbol{n}}\cdot \boldsymbol{\sigma})\sin\abs{\boldsymbol{n}}
\]
For $\boldsymbol{n} = \frac{\pi}{2}(\cos\varphi,\sin\varphi,0)$ we get Eq. (\ref{eq:DC}).

\mfaC{
\subsection{Average Landau Zener}
For $a\sim \mathcal{N}(\mu,\sigma^2)$, the average value of LZ transition probability $\mathcal{P}_\text{LZ}(a) = \text{e}^{-\pi a^2}$, Eq. \eqref{eq:PLZ},
is given by the integral
\begin{equation}
    \langle\mathcal{P}_\text{LZ}\rangle(\mu,\sigma) = \frac{1}{\sqrt{2\pi \sigma^2}}\int_\mathbb{R} \text{d}a\ \text{e}^{-\tfrac{(a-\mu)^2}{2\sigma^2}}\text{e}^{-\pi a^2}.
    \label{eq:PLZavInt}
\end{equation}
Eq. \eqref{eq:PLZavInt} is a standard Gaussian integral which can be calculated by completing the square
\[
    \frac{(a-\mu)^2}{2\sigma^2} + \pi a^2 = \frac{1}{2 (\tfrac{\sigma}{\sqrt{1+2\pi \sigma^2}})^2}\bigg(a-\frac{\mu}{1+2\pi \sigma^2}\bigg)^2
        + \frac{\pi\mu^2}{1+2\pi \sigma^2}.
\]
By shifting the integration parameter $a\rightarrow a+\frac{\mu}{1+2\pi\sigma^2}$ and using the normalization of the Gaussian distribution, $\langle 1\rangle_a = 1$ we get the final formula
\begin{equation}
    \langle\mathcal{P}_\text{LZ}\rangle(\mu,\sigma) = \frac{1}{\sqrt{1+2\pi\sigma^2}}\exp\left( -\frac{\pi \mu^2}{1+2\pi \sigma^2}\right).
    \label{eq:PLZav}
\end{equation}
For $\sigma\rightarrow0$, we recover the LZ formula $\langle\mathcal{P}_\text{LZ}\rangle(\mu,0) = \text{e}^{-\pi \mu^2}$ and for $\sigma\rightarrow\infty$, the leading order term is $\langle\mathcal{P}_\text{LZ}\rangle(\mu,\sigma) \stackrel{\sigma\rightarrow\infty}{\sim} \frac{1}{\sigma\sqrt{2\pi}}$, independently of $\mu$.
}
\subsection{Angle independence}
\label{ap:angle}
In Eq. (\ref{eq:GLZS}), for $a = 0$, the angle parameter $\varphi$ can be factored out.

Define $U_\varphi \equiv \exp(i \frac{\varphi}{2}\sigma_3)$. Expanding the matrix exponential, $U_\varphi = \cos\frac{\varphi}{2}\mathbf{1}+i \sin\frac{\varphi}{2}\sigma_3$ and using the properties of the Pauli matrices and trigonometric identities we obtain
\begin{align}
    U_\varphi^\dagger \sigma_1 U_\varphi
        &= (\cos\tfrac{\varphi}{2}-i \sin\tfrac{\varphi}{2}\sigma_3)\sigma_1(\cos\tfrac{\varphi}{2}+i \sin\tfrac{\varphi}{2}\sigma_3) \nonumber\\
        &= \cos^2\tfrac{\varphi}{2}\sigma_1+i \sin\tfrac{\varphi}{2}\cos\tfrac{\varphi}{2}[\sigma_1,\sigma_3]+\sin\tfrac{\varphi}
            {2}\sigma_3\sigma_1\sigma_3\nonumber\\
        &= (\cos^2\frac{\varphi}{2}-\sin^2\frac{\varphi}{2})\sigma_1+2\sin\frac{\varphi}{2}\cos\frac{\varphi}{2}\sigma_2\nonumber\\
        &=\sigma_1\cos\varphi+\sigma_2\sin\varphi\nonumber\\
        &=\sigma_\varphi \,.
\label{eq:Uphi}
\end{align}
For $a = 0$, the Hamiltonian in Eq. (\ref{eq:GLZS}) becomes $H[\varphi] \equiv -t\sigma_3+f_\text{D}\sigma_\varphi$. 
Combining with Eq. (\ref{eq:Uphi}) and the relation $[U_\varphi,\sigma_3] = 0$ we have
\[
    H[\varphi] = U_\varphi^\dagger H[0]U_\varphi
\]
and thus $\mathcal{U}[\varphi] = U_\varphi^\dagger \mathcal{U}[0] U_\varphi$ with $\mathcal{U}$ the corresponding time evolution matrices. Finally, for the transition probabilities
\begin{align*}
    \mathcal{P}[\varphi]
        &= \abs{\braket{+| \mathcal{U}[\varphi]|+}}^2\\
        &= \abs{\braket{+| U_\varphi^\dagger \mathcal{U}[0] U_\varphi|+}}^2,
            \quad U_\varphi\ket{+} = \text{e}^{i\frac{\varphi}{2}}\ket{+} \\
        &= \abs{\braket{+| \mathcal{U}[0] |+}}^2 \\
        &= \mathcal{P}[0]
\end{align*}
which completes the calculation.

\subsection{$\sigma$ limits}
\label{ap:limits}
Calculation of the two limits in Fig. \ref{fig:6}. We set the notation
\[
    \partial_a \mathcal{P}(a,b) \equiv \mathcal{P}'(a,b),\quad \partial_b \mathcal{P}(a,b) \equiv \dot{\mathcal{P}}(a,b)
\]
and similarly for higher derivatives.

Unpacking the definition of $\langle \mathcal{P}\rangle$ in Eq. (\ref{eq:pbstar}) we have 
\[
    \langle\mathcal{P}\rangle = \frac{1}{\sqrt{2\pi \sigma^2}}\int_\mathbb{R}\text{d}a\ \text{e}^{-(a-\mu)^2/(2\sigma^2)}\mathcal{P}(a,b)
\]

\subsubsection{$\sigma\rightarrow 0$}

Expanding $\mathcal{P}(a,b)$ around $a = \mu$ we get
\[
    \mathcal{P}(a,b) = \mathcal{P}(\mu,b) + \mathcal{P}'(\mu,b)(a-\mu)+\frac{1}{2}\mathcal{P}''(\mu,b)(a-\mu)^2 +\dots
\]
Odd powers vanish under the average due to even-odd symmetry. Using  $\langle (a-\mu)^2\rangle = \sigma^2$
\[
    \langle \mathcal{P}\rangle = \mathcal{P}(\mu,b) + \frac{1}{2}\mathcal{P}''(\mu,b)\sigma^2+\dots
\]
In the last equation we set $b = b^\ast$, Eq. (\ref{eq:bstar}) and expand around $b_0$
\begin{align*}
    \mathcal{P}^\ast\equiv \langle \mathcal{P}(a,b^\ast)\rangle
        &= \mathcal{P}(\mu,b^\ast) + \frac{1}{2}\mathcal{P}''(\mu,b^\ast)\sigma^2+\dots\\
        &= \bigg[\cancel{\mathcal{P}(\mu,b_0)+\dot{\mathcal{P}}(\mu,b_0)(b^\ast-b_0)}+ \frac{1}{2}\ddot{\mathcal{P}}(\mu,b_0)(b^\ast-b_0)^2+\dots\bigg]+
            \frac{1}{2}\mathcal{P}''(\mu,b_0)\sigma^2+\dots \\
        &= \frac{1}{2}\mathcal{P}''(\mu,b_0(\mu))\sigma^2 + \frac{1}{2}\ddot{\mathcal{P}}(\mu,b_0(\mu))b_2^2(\mu)\sigma^4 + \dots\\
        &= \frac{1}{2}\mathcal{P}''(\mu,b_0(\mu))\sigma^2 +\dots
\end{align*}
The terms $\mathcal{P}(\mu,b_0(\mu)),\dot{\mathcal{P}(\mu,b_0(\mu))}$ vanish due to the definition of of $b_0(\mu)$, see Eqs. \eqref{eq:pbstar}, \eqref{eq:CC} and Fig. \ref{fig:1} (right panel).

\subsubsection{$\sigma\rightarrow \infty$}
Let $\xi \equiv 1/\sigma$. We expand $\xi$ around $0$
\begin{align*}
    \langle \mathcal{P}\rangle
        &= \frac{\xi}{\sqrt{2\pi}}\int_\mathbb{R}\text{d}a\ \text{e}^{-\xi^2(a-\mu)^2/2}\mathcal{P}(a,b)\\
        &= \frac{\xi}{\sqrt{2\pi}}\int_\mathbb{R}\text{d}a\ \bigg(1-\frac{1}{2}\xi^2(a-\mu)^2+\dots\bigg)\mathcal{P}(a,b)\\
\end{align*}    
Assuming we can expand $b^\ast$ around $\sigma\rightarrow \infty$ ($\xi\rightarrow 0$)
\[
    b^\ast(\mu,\xi) = b_\infty(\mu) +b_{\infty,2}(\mu)\xi^2 + \dots
\]
\begin{align*}
    \mathcal{P}^\ast \equiv\langle \mathcal{P}(a,b^\ast)\rangle
        &= \frac{\xi}{\sqrt{2\pi}}\int_\mathbb{R}\text{d}a\ \bigg(1-\frac{1}{2}\xi^2(a-\mu)^2+\dots\bigg)\mathcal{P}(a,b^\ast(\mu,\xi))\\
        &= \frac{\xi}{\sqrt{2\pi}}\int_\mathbb{R}\text{d}a\ \bigg(1-\frac{1}{2}\xi^2(a-\mu)^2+\dots\bigg)\mathcal{P}(a,b_\infty(\mu)+\dots)\\
        &= \frac{\xi}{\sqrt{2\pi}}\int_\mathbb{R}\text{d}a\ \bigg(1-\frac{1}{2}\xi^2(a-\mu)^2+\dots\bigg)(\mathcal{P}(a,b_\infty(\mu))+\dots)\\
        &= \frac{\xi}{\sqrt{2\pi}} \int_\mathbb{R} \text{d}a\ \mathcal{P}(a,b_\infty(\mu))+\dots\\
        & \stackrel{\xi\rightarrow 0}{\longrightarrow} 0
\end{align*}

\section{Special Functions}
\label{ap:special}
Properties of the (Euler) Gamma function used in the calculations \cite{abramowitz}

\begin{equation}
    \Gamma(z)\Gamma\left(z+\tfrac{1}{2}\right) = 2^{1-2z}\sqrt{\pi}\Gamma(2z)
    \label{eq:gamma3}
\end{equation}
\begin{equation}
    \abs{\Gamma(\tfrac{1}{2}+bi)}^2 = \frac{\pi}{\cosh\pi b}
    \label{eq:gamma5}
\end{equation}
\begin{equation}
    \abs{\Gamma\left(1+bi\right)}^2 = \frac{\pi b}{\sinh \pi b}
    \label{eq:gamma6}
\end{equation}

\medskip

\section*{References}

\providecommand{\newblock}{}


\begin{thebibliography}{10}
\expandafter\ifx\csname url\endcsname\relax
  \def\url#1{{\tt #1}}\fi
\expandafter\ifx\csname urlprefix\endcsname\relax\def\urlprefix{URL }\fi
\providecommand{\eprint}[2][]{\url{#2}}

\bibitem{vitanov-finite}
Vitanov N~V and Garraway B~M 1996 {\em Phys. Rev. A\/} {\bf 53}(6) 4288--4304
  \urlprefix\url{https://link.aps.org/doi/10.1103/PhysRevA.53.4288}

\bibitem{nori2}
Ivakhnenko O~V, Shevchenko S~N and Nori F 2023 {\em Phys. Rept.\/} {\bf 995}
  1--89

\bibitem{LZ-landau}
Landau L~D 1932 {\em Phys. Z. Sowjetunion\/} {\bf 2}

\bibitem{LZ-zener}
Zener C 1932 {\em Proceedings of the Royal Society of London. Series A,
  Containing Papers of a Mathematical and Physical Character\/} {\bf 137}
  696--702 ISSN 09501207 \urlprefix\url{http://www.jstor.org/stable/96038}

\bibitem{LZ-stuck}
{St\"uckelberg, E C G} 1932 {\em Helv. Phys. Acta\/} {\bf 5} 369
  \urlprefix\url{https://cir.nii.ac.jp/crid/1370289641671472000}

\bibitem{LZ-majorana}
Majorana E 1932 {\em Il Nuovo Cimento\/} {\bf 9} 43?50 ISSN 1827-6121
  \urlprefix\url{http://dx.doi.org/10.1007/BF02960953}

\bibitem{born-fock}
Born M and Fock V 1928 {\em Zeitschrift f\"ur Physik\/} {\bf 51} 165?180 ISSN
  1434-601X \urlprefix\url{http://dx.doi.org/10.1007/BF01343193}

\bibitem{berry-phase}
Berry M~V 1984 {\em Proceedings of the Royal Society of London. A. Mathematical
  and Physical Sciences\/} {\bf 392} 45?57 ISSN 0080-4630
  \urlprefix\url{http://dx.doi.org/10.1098/rspa.1984.0023}

\bibitem{farhi}
Farhi E, Goldstone J, Gutmann S and Sipser M 2000 Quantum computation by
  adiabatic evolution (\textit{Preprint} \eprint{quant-ph/0001106})
  \urlprefix\url{https://arxiv.org/abs/quant-ph/0001106}

\bibitem{Guery}
Gu\'ery-Odelin D, Ruschhaupt A, Kiely A, Torrontegui E, Mart\'{\i}nez-Garaot S
  and Muga J~G 2019 {\em Rev. Mod. Phys.\/} {\bf 91}(4) 045001
  \urlprefix\url{https://link.aps.org/doi/10.1103/RevModPhys.91.045001}

\bibitem{STA-2}
del Campo A and Kim K 2019 {\em New Journal of Physics\/} {\bf 21} 050201
  \urlprefix\url{https://dx.doi.org/10.1088/1367-2630/ab1437}

\bibitem{Dong10}
Dong D and Petersen I 2010 {\em IET Control Theory \& Applications\/} {\bf
  4}(12) 2651--2671
  \urlprefix\url{https://digital-library.theiet.org/doi/abs/10.1049/iet-cta.2009.0508}

\bibitem{DAlessandro21}
D'Alessandro D 2021 {\em Introduction to Quantum Control and Dynamics\/}
  (Chapman and Hall/CRC)
  \urlprefix\url{http://dx.doi.org/10.1201/9781003051268}

\bibitem{Koch22}
Koch C~P, Boscain U, Calarco T, Dirr G, Filipp S, Glaser S~J, Kosloff R,
  Montangero S, Schulte-Herbr\"{u}ggen T, Sugny D and Wilhelm F~K 2022 {\em EPJ
  Quantum Technology\/} {\bf 9} ISSN 2196-0763
  \urlprefix\url{http://dx.doi.org/10.1140/epjqt/s40507-022-00138-x}

\bibitem{Poggi24}
Poggi P~M, De~Chiara G, Campbell S and Kiely A 2024 {\em Phys. Rev. Lett.\/}
  {\bf 132}(19) 193801
  \urlprefix\url{https://link.aps.org/doi/10.1103/PhysRevLett.132.193801}

\bibitem{Weidner24}
Weidner C~A, Reed E~A, Monroe J, Sheller B, O'Neil S, Maas E, Jonckheere E~A,
  Langbein F~C and Schirmer S~G 2024 Robust quantum control in closed and open
  systems: Theory and practice (\textit{Preprint} \eprint{2401.00294})
  \urlprefix\url{https://arxiv.org/abs/2401.00294}

\bibitem{Koch16}
Koch C~P 2016 {\em Journal of Physics: Condensed Matter\/} {\bf 28} 213001
  \urlprefix\url{https://doi.org/10.1088/0953-8984/28/21/213001}

\bibitem{Li09}
Li J~S and Khaneja N 2009 {\em IEEE Transactions on Automatic Control\/} {\bf
  54} 528--536

\bibitem{demirplak1}
Demirplak M and Rice S~A 2003 {\em The Journal of Physical Chemistry A\/} {\bf
  107} 9937--9945 (\textit{Preprint}
  \eprint{https://doi.org/10.1021/jp030708a})
  \urlprefix\url{https://doi.org/10.1021/jp030708a}

\bibitem{demirplak2}
Demirplak M and Rice S~A 2005 {\em The Journal of Physical Chemistry B\/} {\bf
  109} 6838?6844 ISSN 1520-5207
  \urlprefix\url{http://dx.doi.org/10.1021/jp040647w}

\bibitem{berry-trans}
Berry M~V 2009 {\em Journal of Physics A: Mathematical and Theoretical\/} {\bf
  42} 365303 \urlprefix\url{https://dx.doi.org/10.1088/1751-8113/42/36/365303}

\bibitem{wimberger-24}
Petiziol F, Mintert F and Wimberger S 2024 {\em EPL\/} {\bf 145} 15001
  (\textit{Preprint} \eprint{2402.04936})

\bibitem{polk-sels}
Sels D and Polkovnikov A 2017 {\em Proceedings of the National Academy of
  Sciences\/} {\bf 114} E3909--E3916 (\textit{Preprint}
  \eprint{https://www.pnas.org/doi/pdf/10.1073/pnas.1619826114})
  \urlprefix\url{https://www.pnas.org/doi/abs/10.1073/pnas.1619826114}

\bibitem{app1}
Zhang Z, Wang T, Xiang L, Jia Z, Duan P, Cai W, Zhan Z, Zong Z, Wu J, Sun L,
  Yin Y and Guo G 2018 {\em New Journal of Physics\/} {\bf 20} 085001
  \urlprefix\url{https://dx.doi.org/10.1088/1367-2630/aad4e7}

\bibitem{app2}
Veps\"al\"ainen A, Danilin S and Paraoanu G~S 2019 {\em Science Advances\/} {\bf
  5} eaau5999 (\textit{Preprint}
  \eprint{https://www.science.org/doi/pdf/10.1126/sciadv.aau5999})
  \urlprefix\url{https://www.science.org/doi/abs/10.1126/sciadv.aau5999}

\bibitem{app3}
Chen S~Q and Lu H 2025 {\em Phys. Rev. Appl.\/} {\bf 23}(2) 024015
  \urlprefix\url{https://link.aps.org/doi/10.1103/PhysRevApplied.23.024015}

\bibitem{Weitz1994}
Weitz M, Young B~C and Chu S 1994 {\em Phys. Rev. Lett.\/} {\bf 73}(19)
  2563--2566
  \urlprefix\url{https://link.aps.org/doi/10.1103/PhysRevLett.73.2563}

\bibitem{Bateman2007}
Bateman J and Freegarde T 2007 {\em Phys. Rev. A\/} {\bf 76}(1) 013416
  \urlprefix\url{https://link.aps.org/doi/10.1103/PhysRevA.76.013416}

\bibitem{Wubs2005}
Wubs M, Saito K, Kohler S, Kayanuma Y and H\"anggi P 2005 {\em New Journal of
  Physics\/} {\bf 7} 218
  \urlprefix\url{https://dx.doi.org/10.1088/1367-2630/7/1/218}

\bibitem{exper1}
Villazon T, Claeys P~W, Polkovnikov A and Chandran A 2021 {\em Phys. Rev. B\/}
  {\bf 103}(7) 075118
  \urlprefix\url{https://link.aps.org/doi/10.1103/PhysRevB.103.075118}

\bibitem{exper2}
Sompet P, Szigeti S~S, Schwartz E, Bradley A~S and Andersen M~F 2019 {\em
  Nature Communications\/} {\bf 10} ISSN 2041-1723
  \urlprefix\url{http://dx.doi.org/10.1038/s41467-019-09420-6}

\bibitem{LZC-1}
Wittig C 2005 {\em J. Phys. Chem. B\/} {\bf 109} 8428--8430

\bibitem{LZC-2}
Chichinin A 2013 {\em The Journal of Physical Chemistry B\/} {\bf 117}
  6018--6018

\bibitem{LZC-3}
Ho L~T~A and Chibotaru L~F 2014 {\em Phys. Chem. Chem. Phys.\/} {\bf 16}(15)
  6942--6945

\bibitem{LZD-kayanuma}
Kayanuma Y 1984 {\em Journal of the Physical Society of Japan\/} {\bf 53}
  108?117 ISSN 1347-4073
  \urlprefix\url{http://dx.doi.org/10.1143/JPSJ.53.108}

\bibitem{LZD-rojo}
Rojo A~G 2010 Matrix exponential solution of the {Landau-Zener} problem
  (\textit{Preprint} \eprint{1004.2914})
  \urlprefix\url{https://arxiv.org/abs/1004.2914}

\bibitem{LZother}
Glasbrenner E~P and Schleich W~P 2023 {\em Journal of Physics B: Atomic,
  Molecular and Optical Physics\/} {\bf 56} 104001

\bibitem{qmbook-Sakurai}
Sakurai J~J and Napolitano J 2020 {\em Modern Quantum Mechanics\/} 3rd ed
  (Cambridge University Press)

\bibitem{wimberger-18}
Petiziol F, Dive B, Mintert F and Wimberger S 2018 {\em Phys. Rev. A\/} {\bf
  98}(4) 043436
  \urlprefix\url{https://link.aps.org/doi/10.1103/PhysRevA.98.043436}

\bibitem{qmbook-Messiah}
Messiah A 1961 {\em Quantum Mechanics Volume II\/} (Elsevier Science B.V.)

\bibitem{polk-sels2}
Claeys P~W, Pandey M, Sels D and Polkovnikov A 2019 {\em Phys. Rev. Lett.\/}
  {\bf 123}(9) 090602
  \urlprefix\url{https://link.aps.org/doi/10.1103/PhysRevLett.123.090602}

\bibitem{qmbook-NielsenChuang}
Nielsen M~A and Chuang I~L 2010 {\em Quantum Computation and Quantum
  Information: 10th Anniversary Edition\/} (Cambridge University Press)

\bibitem{abramowitz}
Abramowitz M and Stegun I~A 1964 {\em Handbook of Mathematical Functions with
  Formulas, Graphs, and Mathematical Tables\/} ninth dover printing, tenth gpo
  printing ed (New York: Dover)

\bibitem{vitanov-time}
Vitanov N~V 1999 {\em Phys. Rev. A\/} {\bf 59}(2) 988--994
  \urlprefix\url{https://link.aps.org/doi/10.1103/PhysRevA.59.988}

\bibitem{nlRC}
Roland J and Cerf N~J 2002 {\em Phys. Rev. A\/} {\bf 65}(4) 042308
  \urlprefix\url{https://link.aps.org/doi/10.1103/PhysRevA.65.042308}

\bibitem{Lidar2010}
Rezakhani A~T, Pimachev A~K and Lidar D~A 2010 {\em Phys. Rev. A\/} {\bf 82}(5)
  052305 \urlprefix\url{https://link.aps.org/doi/10.1103/PhysRevA.82.052305}

\bibitem{nl-tan}
Rezakhani A~T, Kuo W~J, Hamma A, Lidar D~A and Zanardi P 2009 {\em Physical
  Review Letters\/} {\bf 103} ISSN 1079-7114
  \urlprefix\url{http://dx.doi.org/10.1103/PhysRevLett.103.080502}

\bibitem{bason1}
Bason M~G, Viteau M, Malossi N, Huillery P, Arimondo E, Ciampini D, Fazio R,
  Giovannetti V, Mannella R and Morsch O 2012 {\em Nature Physics\/} {\bf 8}
  147?152 ISSN 1745-2481 \urlprefix\url{http://dx.doi.org/10.1038/nphys2170}

\bibitem{Polkovnikov2016}
Tomka M, Souza T, Rosenberg S and Polkovnikov A 2016 Geodesic paths for quantum
  many-body systems (\textit{Preprint} \eprint{1606.05890})
  \urlprefix\url{https://arxiv.org/abs/1606.05890}

\bibitem{wimberger-19}
Petiziol F, Dive B, Carretta S, Mannella R, Mintert F and Wimberger S 2019 {\em
  Phys. Rev. A\/} {\bf 99}(4) 042315
  \urlprefix\url{https://link.aps.org/doi/10.1103/PhysRevA.99.042315}

\bibitem{polk-geometry}
Kolodrubetz M, Sels D, Mehta P and Polkovnikov A 2017 {\em Physics Reports\/}
  {\bf 697} 1--87 ISSN 0370-1573
  \urlprefix\url{https://www.sciencedirect.com/science/article/pii/S0370157317301989}

\bibitem{Dengis2025}
Dengis S, Wimberger S and Schlagheck P 2025 {\em Phys. Rev. A\/} {\bf 111}(3)
  L031301 \urlprefix\url{https://link.aps.org/doi/10.1103/PhysRevA.111.L031301}

\bibitem{Dengis2025-2}
Dengis S, Wimberger S and Schlagheck P 2025 {\em Phys. Rev. A\/} {\bf 112}(4)
  042610 \urlprefix\url{https://link.aps.org/doi/10.1103/qxqg-qnnq}

\bibitem{Romanato2025}
Romanato L, Eshaqi-Sani N, Lepori L, Kirova T, Arimondo E and Wimberger S 2025
  {\em Phys. Rev. A\/} {\bf 112}(5) 052812
  \urlprefix\url{https://link.aps.org/doi/10.1103/pst6-r584}

\bibitem{vitanov-nls}
Vitanov N~V and Suominen K~A 1999 {\em Phys. Rev. A\/} {\bf 59}(6) 4580--4588
  \urlprefix\url{https://link.aps.org/doi/10.1103/PhysRevA.59.4580}

\bibitem{LZ-nls}
Ashhab S, Ilinskaya O~A and Shevchenko S~N 2022 {\em Phys. Rev. A\/} {\bf
  106}(6) 062613
  \urlprefix\url{https://link.aps.org/doi/10.1103/PhysRevA.106.062613}

\bibitem{Suresh2023}
Suresh A, Varma V, Batra P and Mahesh T~S 2023 {\em The Journal of Chemical
  Physics\/} {\bf 159} 024202 ISSN 0021-9606
  \urlprefix\url{https://doi.org/10.1063/5.0159448}

\bibitem{kayanuma-offdiagonal}
Kayanuma Y 1985 {\em Journal of the Physical Society of Japan\/} {\bf 54}
  2037--2046

\bibitem{Ao1989}
Ao P and Rammer J 1989 {\em Phys. Rev. Lett.\/} {\bf 62}(25) 3004--3007
  \urlprefix\url{https://link.aps.org/doi/10.1103/PhysRevLett.62.3004}

\bibitem{Wubs2006}
Wubs M, Saito K, Kohler S, H\"anggi P and Kayanuma Y 2006 {\em Phys. Rev.
  Lett.\/} {\bf 97}(20) 200404
  \urlprefix\url{https://link.aps.org/doi/10.1103/PhysRevLett.97.200404}

\bibitem{Kayanuma-diss}
Kayanuma Y and Nakayama H 1998 {\em Phys. Rev. B\/} {\bf 57}(20) 13099--13112
  \urlprefix\url{https://link.aps.org/doi/10.1103/PhysRevB.57.13099}

\bibitem{Huang2018}
Huang Z and Zhao Y 2018 {\em Phys. Rev. A\/} {\bf 97}(1) 013803
  \urlprefix\url{https://link.aps.org/doi/10.1103/PhysRevA.97.013803}

\bibitem{Pokrovsky-fastNoise}
Pokrovsky V~L and Sinitsyn N~A 2003 {\em Phys. Rev. B\/} {\bf 67}(14) 144303
  \urlprefix\url{https://link.aps.org/doi/10.1103/PhysRevB.67.144303}

\bibitem{Kiely2021}
Kiely A 2021 {\em Europhysics Letters\/} {\bf 134} 10001
  \urlprefix\url{https://doi.org/10.1209/0295-5075/134/10001}

\bibitem{CDopen-Vacanti}
Vacanti G, Fazio R, Montangero S, Palma G~M, Paternostro M and Vedral V 2014
  {\em New Journal of Physics\/} {\bf 16} 053017
  \urlprefix\url{https://doi.org/10.1088/1367-2630/16/5/053017}

\bibitem{CDopen-Query}
Impens F and Gu{\'e}ry-Odelin D 2019 {\em Sci. Rep.\/} {\bf 9} 4048

\bibitem{CDopen-Alipour}
Alipour S, Chenu A, Rezakhani A~T and del Campo A 2020 {\em {Quantum}\/} {\bf
  4} 336 ISSN 2521-327X
  \urlprefix\url{https://doi.org/10.22331/q-2020-09-28-336}

\bibitem{CDopne-Santos}
Santos A~C and Sarandy M~S 2021 {\em Phys. Rev. A\/} {\bf 104}(6) 062421
  \urlprefix\url{https://link.aps.org/doi/10.1103/PhysRevA.104.062421}

\bibitem{whittaker}
Whittaker E~T and Watson G~N 1996 {\em A Course of Modern Analysis\/}
  (Cambridge University Press) ISBN 9780511608759
  \urlprefix\url{http://dx.doi.org/10.1017/CBO9780511608759}

\bibitem{asymp}
Vidunas R and Temme N~M 2002  (\textit{Preprint} \eprint{math/0205045})
  \urlprefix\url{https://arxiv.org/abs/math/0205045}

\bibitem{vitanov-book}
Vitanov N~V 2010 {\em Quantum Transitions\/} (St. Kliment Ohridski University
  Press)

\end{thebibliography}

\end{document}